\documentclass[superscriptaddress,twocolumn,prb]{revtex4-2}
\usepackage{bm}
\usepackage{amssymb}
\usepackage{lipsum}
\usepackage{amsmath}
\usepackage{braket}
\usepackage{float}
\usepackage{stfloats}
\usepackage{multirow}
\usepackage{changes}
\usepackage{xcolor}
\usepackage{graphicx}
\usepackage[colorlinks=True,linkcolor=blue,anchorcolor=red,citecolor=blue,CJKbookmarks=true,urlcolor=blue]{hyperref}

\begin{document}
\title{Skin effect in Non-Hermitian systems with spin}
\author{Wenna Zhang}
\affiliation{School of Physics, Central South University, Changsha, 410083, China}
\author{Yutao Hu}
\affiliation{Shenzhen Institute for Quantum Science and Engineering, Southern university of science and technology, Shenzhen, 518055, China}
\affiliation{School of Physics, Central South University, Changsha, 410083, China}
\author{Hongyi Zhang}
\affiliation{School of Physics, Central South University, Changsha, 410083, China}
\author{Xiang Liu}
\affiliation{School of Physics, Central South University, Changsha, 410083, China}
\author{Yuecheng Shen} 
\affiliation{State Key Laboratory of Precision Spectroscopy, East China Normal University, Shanghai, 200241, China}
\author{Georgios Veronis}
\affiliation{School of Electrical Engineering and Computer Science, Louisiana State University, Baton Rouge, Louisiana 70803, USA}	
\affiliation{Center for Computation and Technology, Louisiana State University, Baton Rouge, Louisiana 70803, USA}
\author{Andrea Al\`u}
\affiliation{Photonics Initiative, Advanced Science Research Center, City University of New York, New York, NY 10031, USA}	
\affiliation{Physics Program, Graduate Center, City University of New York, New York, NY 10026, USA}
\author{Yin Huang} \email{yhuan15@csu.edu.cn}
\affiliation{School of Physics, Central South University, Changsha, 410083, China}
\author{Wenchen Luo}\email{luo.wenchen@csu.edu.cn}
\affiliation{School of Physics, Central South University, Changsha, 410083, China}
\date{\today}

\begin{abstract}
The skin effect, where bulk modes collapse into boundary modes, is 
a key phenomenon in topological non-Hermitian systems, has been predominantly studied in spinless systems. 
Recent studies illustrate the magnetic suppression of the first-order skin effect while ignoring spin. 
However, the physical significance of a magnetic field in non-Hermitian skin effect with spin remains elusive. Here, we systematically explore non-Hermitian spinful systems based on generalized Hatano-Nelson models with SU(2) gauge potential fields. In an open one-dimensional lattice, the spin-up and spin-down states can be uniquely separated and localized at the two boundaries without magnetic field. When an external magnetic field is applied, the skin effect exhibits a smooth transition from bidirectional to unidirectional. Remarkably, we demonstrate that the first-order skin effect can be anomalously induced by a magnetic field in a topologically trivial non-Hermitian spinful system without any skin effect at zero field. 
The direction of such magnetically induced skin modes can be controlled by simply changing the amplitude and polarity of the magnetic field. In addition, we demonstrate a transition between non-Bloch $\mathcal{PT}$ and anti-$\mathcal{PT}$ symmetries in the system, and uncover the spin-dependent mechanism of non-Bloch $\mathcal{PT}$ symmetry. Our results pave the way for the investigation of non-Hermitian skin effect with spin degrees of freedom.

\end{abstract}

\maketitle

\textit{Introduction.}\textemdash 
Non-Hermitian physics has flourished over the past few decades, revealing numerous unique phenomena \cite{peng2014parity,ramezani2010unidirectional,chang2014parity,chen2016topological,zhou2018observation,shen2018quantum,papaj2019nodal,yoshida2018non}. One of these intriguing phenomena is the non-Hermitian skin effect (NHSE) \cite{yao2018edge, PhysRevX.9.041015,zhang2022review, slager, Zhang2021Correspondence,kunst2018biorthogonal,Li2022Gain}, characterized by the localization of eigenstates at boundaries. This edge effect, which leads to the breakdown of bulk-boundary correspondence associated with the point gap topology, has widespread applications across various fields of physics \cite{weidemann2020topological,wang2021generating,xiao2020non,zhang2021observation,zhang2021acoustic,gu2022transient,wang2022non,li2020critical,helbig2020generalized,zhu2022anomalous,franca2022non,Longhi2022Self}. Thus far, studies of the NHSE have mostly focused on spinless systems. Recent studies demonstrated that the first-order skin effect (FOSE) in spinless lattices can be significantly suppressed by applying an external magnetic field \cite{lu2021magnetic,Shao2022Cyclotron}. Electrons undergo cyclotron motion and are localized in the bulk of the system under an external magnetic field, leading to a suppression of FOSE under suitable conditions. It has also been reported that external magnetic fields can enhance the second-order skin effect in spinless systems, while the FOSE is still suppressed \cite{Li2023enhancement}. 
While in topological insulators the internal spin degrees of freedom play a fundamental role \cite{asboth2016short}, the role of spin has been sparsely studied and its physical significance in NHSE remains elusive.

In the presence of non-Hermicity, it is possible to have entirely real spectra under open boundary conditions (OBC), known as non-Bloch $\mathcal{PT}$ symmetry \cite{Hu2024geometric, Xiao2021observation}. The phase transition among different conventional Bloch $\mathcal{PT}$ symmetric phases has been realized \cite{Hu2023anti}. However, the realization of phase transitions among different non-Bloch $\mathcal{PT}$ symmetric phases has yet to be demonstrated. Whereas the geometric mechanism of non-Bloch $\mathcal{PT}$ symmetry and its breaking have been studied \cite{Hu2024geometric}, the spin dependent origin of non-Bloch $\mathcal{PT}$ symmetry has not been investigated yet.

To explore the physics of spinful non-Hermitian lattices, we introduce SU(2) gauge potentials into a one-dimensional (1D) model. We achieve a bidirectional skin effect, typically observed in spinless models with next-nearest-neighbors hopping \cite{wang2021generating}, accompanied by spin separation, in the absence of magnetic field. 
We demonstrate a transition from bidirectional to unidirectional skin effect in the presence of magnetic field. 
Remarkably, we find that a topologically trivial non-Hermitian spinful system without any skin effect in the absence of magnetic field can exhibit a FOSE in the presence of magnetic field.
Such an anomalous phenomenon is due to the emergence of nonreciprocal effective hopping strengths induced by the external magnetic field. The direction of such a magnetically-induced skin effect can also be directly controlled by Zeeman coupling. We further show a transition between non-Bloch $\mathcal{PT}$ and anti-$\mathcal{PT}$ symmetries in our system. The analytical criterion for a transition between them is obtained using perturbation theory. More importantly, we uncover the unique spin-dependent signature of non-Bloch $\mathcal{PT}$ symmetry.

We start from the Hatano-Nelson (HN) model \cite{Hatano1996localization} generalized to include the spin degree of freedom, where two SU(2) phases are embedded in the hopping.  
In the presence of a magnetic field $\boldsymbol{\delta}=(\delta_x,\delta_y,\delta_z)$, the Hamiltonian reads
\begin{equation} \label{modelh}
H_\mu=\sum_{i} c_{i}^{\dag}t_{1}e^{i\theta \mathbf{\sigma_{\mu}}}c_{i+1}
+c_{i+1}^{\dag}t_{2}e^{i\phi \mathbf{\sigma_{\mu}}}c_{i}+c_{i}^{\dag} (\boldsymbol{\delta \cdot \sigma})c_{i},
\end{equation}
where $\mu=x,y,z$ and $\boldsymbol{\sigma}=(\sigma_x, \sigma_y, \sigma_z)$ are Pauli matrices, $c_{i}= \left(\begin{array}{cc}  c_{i,+} & c_{i,-} \end{array}\right)^\intercal$ and $c_{i}^{\dag}$ are the annihilation and creation operators at site $i$ in the basis of $\sigma_z$, and $t_{1(2)}$ is the leftward (rightward) hopping amplitude. The corresponding hopping phases $\theta$ and $\phi$ distinguishing hoppings between different spins, could be complex or real and possibly lead to emergent skin effects associated with spins. Without loss of generality, we choose $\mu=z$, and the gauge can be eliminated by carefully choosing the direction of the magnetic field in the 1D system. To understand the role of SU(2) gauge fields in the NHSE, we set for simplicity $t_{1,2}=1$.

\begin{figure}[htbp]
\centering
\includegraphics[width=0.9\columnwidth, trim=0 2 0 5, clip]{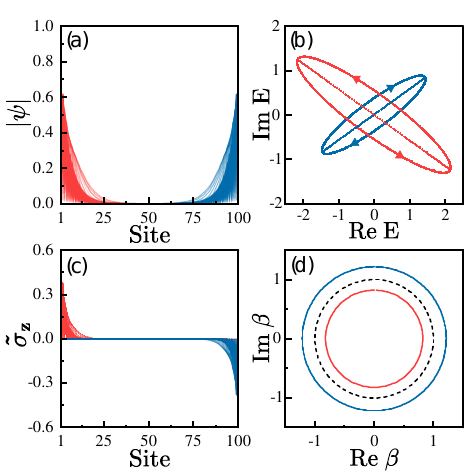}
\caption{
(a) The bidirectional skin effect in the absence of magnetic field, where $\theta=\frac{i \pi}{8}, \phi=\frac{\pi}{3}, \delta=0$ in the model Hamiltonian (Eq. \ref{modelh}). The red and blue curves represent left and right skin modes, respectively. (b) The energy spectra of the system in OBC are surrounded by the energy curves in PBC with winding numbers $\pm1$. (c) The spin separation is associated with the skin effect. (d) The unit circle is located between the GBZs, which also indicates the bidirectional skin effect.}
\label{fig1}
\end{figure}

\textit{Bidirectional skin effect with spin separation.}\textemdash The hopping phases $\theta, \phi$ explicitly contribute to non-Hermiticity. If $\theta$ is purely imaginary and $\phi$ is purely real, the spin-up and spin-down states are separated and skinned to opposite boundaries at zero magnetic field [Fig. \ref{fig1}(a)]. 
To elucidate this, we calculate the eigenenergies of the Hamiltonian under periodic boundary condition (PBC) and OBC and find 
$
E_{\pm}(k)= e^{ ik}e^{\pm i\theta } + e^{-ik}e^{\pm i\phi}, 
$
and
$
E_{\pm}(n)=2\sqrt{  e^{\pm i(\theta+\phi)}} \cos(\frac{n\pi}{N+1}),
$
respectively,
where $k \in [0,2 \pi)$ and $n=1,\ldots N$, with lattice number $N$ in the 1D chain [see Supplemental Material (SM) Sec. I \cite{sm}]. The spectra of spin-up and spin-down states are independent and lie on two distinct Riemann surfaces. 
These two PBC energy bands form two independent closed elliptical loops, each surrounding one of the two linear OBC spectra on the complex plane. On the respective Riemann surfaces [Fig. \ref{fig1}(b)], the OBC eigenstates enclosed by the curve with winding number $\omega = -1$ skin to the right, while the OBC eigenstates with winding number $\omega = 1$ skin to the left. The winding number is defined by $\omega \equiv (1/2\pi)\int_{0}^{2\pi} \partial_{k}\text{arg}[E(k)-E_{b}]\text{d}k$ \cite{PhysRevX.9.041015,Nakamura2024Bulk} with $E_b$ being a reference energy. Here $\omega = \pm 1 $ represent the clockwise and counterclockwise counters of the PBC spectra, respectively.

We define the spin polarization of the $j-$th right (or left) eigenstate $| \psi_j\rangle$ in the lattice at the $i-$th site, 
$\tilde{\sigma}_{j,\mu}^{(i)} \equiv \langle \psi^i_j| \sigma_\mu| \psi^i_j\rangle$, where $| \psi_j^i \rangle$ is a right  
eigenstate at the $i-$th site. By introducing these quantities, the skin effect can be further classified by spin field. Fig. \ref{fig1}(c) shows that the $z$ component spin $\tilde{\sigma}_z$ is localized at the two opposite boundaries, with  opposite signs on different sides, $\tilde{\sigma}_{j,z}^{(i)}=-\tilde{\sigma}_{j',z}^{(N+1-i)}$. This confirms the bidirectional skin effect with spin separation. Note that the other spin components $\tilde{\sigma}_{x,y}$ vanish. The bidirectional skin effect supports a non-Bloch generalized Brillouin zone (GBZ) [Fig. \ref{fig1}(d)] which is abstracted from the non-Bloch Hamiltonian $H(\beta)=H(k)|_{e^{ik}\to\beta}$ \cite{Yokomizo2019non,Hu2024geometric}. In addition, an intuitive way to understand such a bidirectional skin effect with spin separation is to rewrite the Hamiltonian in the form of two copies of the HN model (see SM Sec. II \cite{sm}).

\begin{figure}[htbp]
\centering
\includegraphics[width=0.96\linewidth, trim=0 7 0 7, clip]{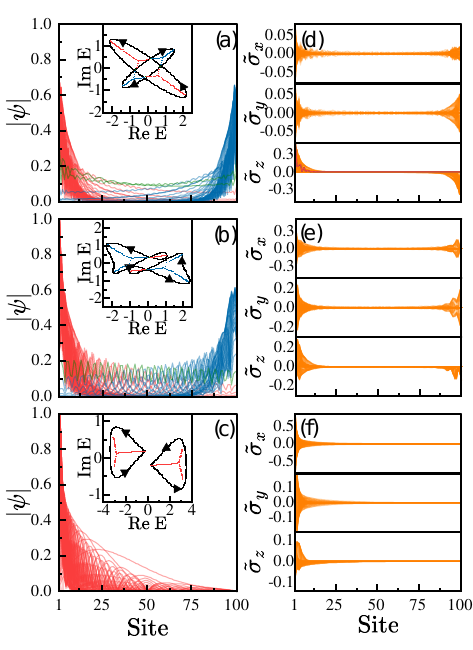}
\caption{
(a) to (c) Evolution of bidirectional to unidirectional skin effect with increasing magnetic field amplitude due to energy spectra topology transition. The system parameters are $\theta=\frac{i \pi}{8}, \phi=\frac{\pi}{3}$, and $\delta_x=0.1,0.8,2$ for (a), (b), (c), respectively. (d) to (f) The spin components $\tilde{\sigma}_{x,y}$ arise with symmetry to $0$. All spin components move to the left as $\delta_x$ increases. The parameters in (d), (e), (f) are the same as in (a), (b), (c), respectively.}
\label{fig2}
\end{figure}
 
\textit{Transition from bidirectional to unidirectional skin effect.}\textemdash A magnetic field directly couples to a spinful system and is easy to manipulate \cite{huang2016experimental,asboth2016short}. 
Here, we need to constrain the magnetic field in any direction other than the $z$ axis. Otherwise the NHSE does not qualitatively change due to $[\sigma_z,H_z]=0$. For simplicity, the magnetic field is assumed to be along the $x$ axis, $\boldsymbol{\delta}=(\delta_x,0,0)$. Its gauge can be chosen to vanish in the 1D system. The Zeeman coupling does not commute to the Hamiltonian, and can effectively manipulate the system.

As shown in Fig. \ref{fig2}, turning on the magnetic field gradually polarizes the symmetric bidirectional skin effect towards one boundary. In the presence of the magnetic field, the two Riemann surfaces corresponding to the two spins are no longer independent and collapse into one due to the interaction between the spin-up and spin-down states. When the external magnetic field is weak, even though the PBC spectra remain almost unchanged, the OBC energy spectra significantly change [Fig. \ref{fig2}(a)]. Since one of the two closed loops for the PBC spectra travels around the origin clockwise while the other one travels counterclockwise, the overall winding number around any interior point in the overlapping region becomes zero. To maintain the skin effect, most of the modes originally located in the overlapping region at zero field, escape the overlapping region and enter the area enclosed with winding number $\omega \neq 0$. 

\begin{figure}[htbp]
\centering
\includegraphics[width=0.96\linewidth, trim=0 3 0 4, clip]{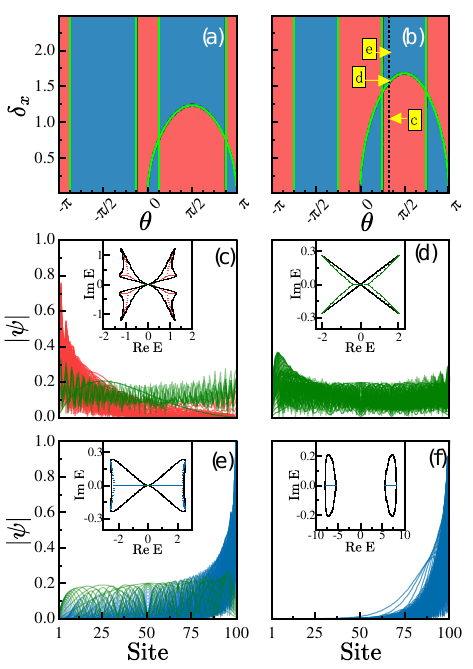}
\caption{(a), (b) Phase diagrams of the non-Hermitian spinful model for $\phi=\frac{7\pi}{8}$
and $\phi=\frac{\pi}{4}$, respectively, featuring topological structures with winding number $\omega$ = 1 (blue region) and -1 (red region). (c) to (f) Evolution of the skin effect with increasing magnetic field amplitude along the dashed line in Fig. 3(b). The system parameters are $\theta=\frac{\pi}{3}$, $\phi=\frac{\pi}{4}$, and $\delta_x=$1, 1.57, 2, and 7 for (c) to (f), respectively.
}
\label{fig3}
\end{figure}

As the magnetic field amplitude increases, the two closed loops gradually deform, and at a critical magnetic field amplitude, the previous two energy spectra recombine to form two new energy spectra symmetric about the origin on the complex plane [Fig. \ref{fig2}(b)]. As the magnetic field amplitude further increases, the PBC spectra become two separate rings with identical winding number [Fig. \ref{fig2}(c)]. Consequently, the bidirectional skin effect becomes a unidirectional skin effect. The final skin direction 
obeys the following rule. If $\left|\cosh(i\theta)\right| > \left|\cosh(i\phi)\right|$, 
leftward hoppings dominate with $\omega=1$ for energy bands in PBC and
the states skin toward the left boundary. Otherwise, the states skin toward the right boundary (see SM Sec. III. A \cite{sm}). 
Fig. \ref{fig2}(d) also shows that positive and negative $z$ component of the spin polarization, $\tilde{\sigma}_z$, are localized at opposite boundaries, when the external magnetic field is weak. As the magnetic field amplitude increases, states with negative $\tilde{\sigma}_z$ gradually reverse to the other side, indicating that the spin separation collapses and both spin-up and spin-down states aggregate to the same side [Figs. \ref{fig2}(e) and \ref{fig2}(f)]. Interestingly, we observe that $\tilde{\sigma}_x$ and $\tilde{\sigma}_y$ arise and are always symmetric about the transverse coordinate, once the magnetic field is applied [Figs. \ref{fig2}(d)-\ref{fig2}(f)] (see SM Sec. IV \cite{sm}). A special case occurs when $|\theta|$ is so large that the bidirectional skin effect is fragile \cite{okuma2019topological,guo2021exact}, and the minimum magnetic field required for the bidirectional to unidirectional skin effect transition is related to the lattice length (see SM Sec. V \cite{sm}).
	
\textit{FOSE induced by magnetic fields.}\textemdash 
In numerous systems, such as in Landau quantization, theories ignoring the spin degree of freedom work well. 
However, the FOSE suppression by an external magnetic field in a spinless system does not occur in our spinful system when both $\theta$ and $\phi$ are real. No skin effect exists in the absence of a magnetic field. The system is essentially topologically trivial, represented by the zero winding number of the energy spectra in PBC forming two straight lines. 

Here, applying a magnetic field does not induce the cyclotron-like motion in a 1D system, but the associated Zeeman coupling is able to change the topological properties of the system.
Figs. \ref{fig3}(a) and \ref{fig3}(b) show the winding number as a function of the gauge field $\theta$ and magnetic field $\mathbf{\delta}$ under different gauge fields $\phi$. Compared to the case without magnetic field, a complex winding of the band structure with point-gap topology is achieved when a magnetic field is applied [Fig. \ref{fig3}(c)]. In particular, the enclosed OBC states all aggregate to one side of the lattice due to the fact that the energy curves in PBC rotate in the same direction. This unexpected feature indicates that the magnetic field alone can induce the FOSE in topologically trivial systems without any skin effect, in stark contrast to spinless cases \cite{lu2021magnetic,Shao2022Cyclotron,Li2023enhancement}. Asymptotic analysis reveals that this unique behavior stems from the magnetic field inducing nonreciprocal effective hopping strengths (see SM Sec. III. B \cite{sm}).

\begin{figure}[h]
\centering
\includegraphics[width=0.98\linewidth]{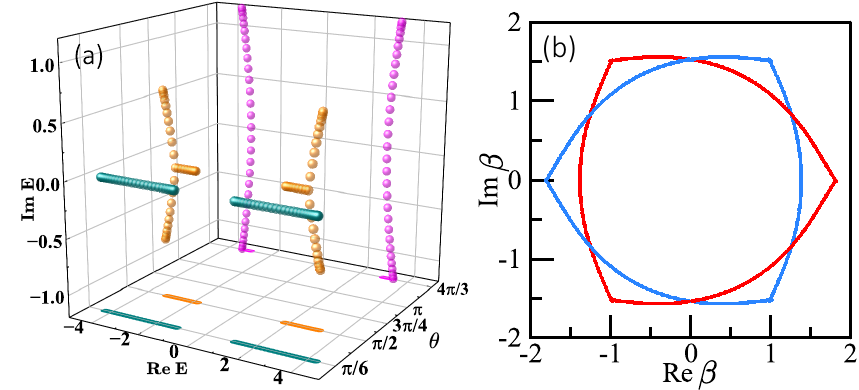}
\caption{(a) Band diagrams with $\delta_x$=7, $\theta=\frac{\pi}{3}$, and $\phi=\frac{\pi}{4}$ (canyon), $\delta_x=7$, $\theta=\frac{\pi}{2}$, and $\phi=\frac{\pi}{4}$ (yellow), and $\delta_x=7, \theta=\frac{2\pi}{3}$, and $\phi=\frac{\pi}{4}$ (magenta) in the $\mathcal{PT}$ symmetric phase, transition point, and anti-$\mathcal{PT}$ symmetric phase, respectively. (b) GBZ at the transition point with the same parameters $\delta_x=7$, $\theta=\frac{\pi}{2}$, but $\phi=\pi$ for simplicity.}
\label{fig4}
\end{figure}

As the magnetic field increases, the winding number changes sign across the curve $\delta_x=2\sqrt{\sin\theta \sin \phi}$ [Figs. \ref{fig3}(a) and \ref{fig3}(b)], which implies a reverse transition of the skin effect. Notably, when $\delta_x=2\sqrt{\sin\theta \sin \phi}$ [green curves in Figs. \ref{fig3}(a) and \ref{fig3}(b)], the PBC spectra again collapse into two straight lines (see SM Sec. III. C \cite{sm}) to reset their winding number to zero, and all modes revert to being extended [Fig. \ref{fig3}(d)]. When $\delta_x > 2\sqrt{\sin \theta \sin \phi} $, the PBC energy spectra again form an enclosed loop but with the opposite winding number, and thus, the OBC states gradually aggregate to the other side [Figs. \ref{fig3}(e) and \ref{fig3}(f)]. This implies that the direction of the magnetically-induced skin effect can be controlled by simply adjusting the amplitude of the magnetic field. We note that when $\sin\theta \sin \phi <0$ this transition will not happen unless $\delta_x$ is complex. We additionally observe that the winding number changes sign across the lines $\phi=\pm \theta$ and $\phi=\pm \theta \mp \pi$ [green lines in Figs. \ref{fig3}(a) and \ref{fig3}(b)]. In addition, the PBC spectra collapse into arcs and all OBC modes become extended as we cross these lines (see SM Sec. III. C \cite{sm}).

\textit{Transition from non-Bloch $\mathcal{PT}$ to anti-$\mathcal{PT}$ symmetric phases.}\textemdash Interestingly, under a sufficiently strong magnetic field, our system exhibits a transition between non-Bloch $\mathcal{PT}$ symmetric and anti-$\mathcal{PT}$ symmetric phases, depending on the hopping phases $\theta$ and $\phi$ [Fig. \ref{fig4}(a)].
A straightforward perturbation calculation (see SM Sec. VI. A \cite{sm}), treating the Hermitian Zeeman coupling as the unperturbed term and the hopping terms as the perturbation, allows one to explicitly classify the symmetries which determine whether the eigenenergies are real or complex. The unperturbed system is degenerate with only two eigenenergies $\pm \delta_x$. The phase of the system is determined as follows: if $\Delta=\cos \theta \cos \phi >0$, all OBC energies become strictly real, since the corrected energies are real-valued functions of the real variable $\sqrt{\Delta}$; if $\Delta<0$, the first-order perturbation contributes only to the imaginary part of the OBC energies, exhibiting strict non-Bloch anti-$\mathcal{PT}$ symmetry. However, higher-order corrections contribute to the real part of the eigenenergies leading to non-Bloch anti-$\mathcal{PT}$ symmetry with complex conjugated OBC spectra [Fig. \ref{fig4}(a)]. A larger Hermitian Zeeman coupling is equivalent to smaller non-Hermiticity. The system therefore tends to be in $\mathcal{PT}$ or anti-$\mathcal{PT}$ symmetric phases under a sufficiently large magnetic field (see SM Sec. VI. B \cite{sm}).

$\Delta =0$ corresponds to the transition point between the $\mathcal{PT}$ and anti-$\mathcal{PT}$ symmetric phases, where a branch of OBC eigenvalues are real while others are complex. 
The formation of the cusps on the GBZ, which indicates that the OBC spectra switch from real to complex, is the geometric origin of non-Bloch $\mathcal{PT}$ symmetry breaking \cite{Hu2024geometric}. Fig. \ref{fig4}(b) shows that, if there are cusps on the GBZ, a phase transition point between non-Bloch $\mathcal{PT}$ and anti-$\mathcal{PT}$ symmetric phases may occur. In contrast, the GBZs for both non-Bloch $\mathcal{PT}$ and anti-$\mathcal{PT}$ symmetric phases are smooth (see SM Sec. VI. C \cite{sm}). However, this also implies that the non-Bloch $\mathcal{PT}$ and anti-$\mathcal{PT}$ symmetries have the same geometric mechanism. The following question then naturally arises: What else is the distinguishing mechanism of non-Bloch $\mathcal{PT}$ symmetry? The answer is the vanishing of $\tilde{\sigma}_z$ ($\tilde{\sigma}_z=0$), as illustrated below. 
\begin{figure}[hbtp]
\centering
\includegraphics[width=0.95\linewidth]{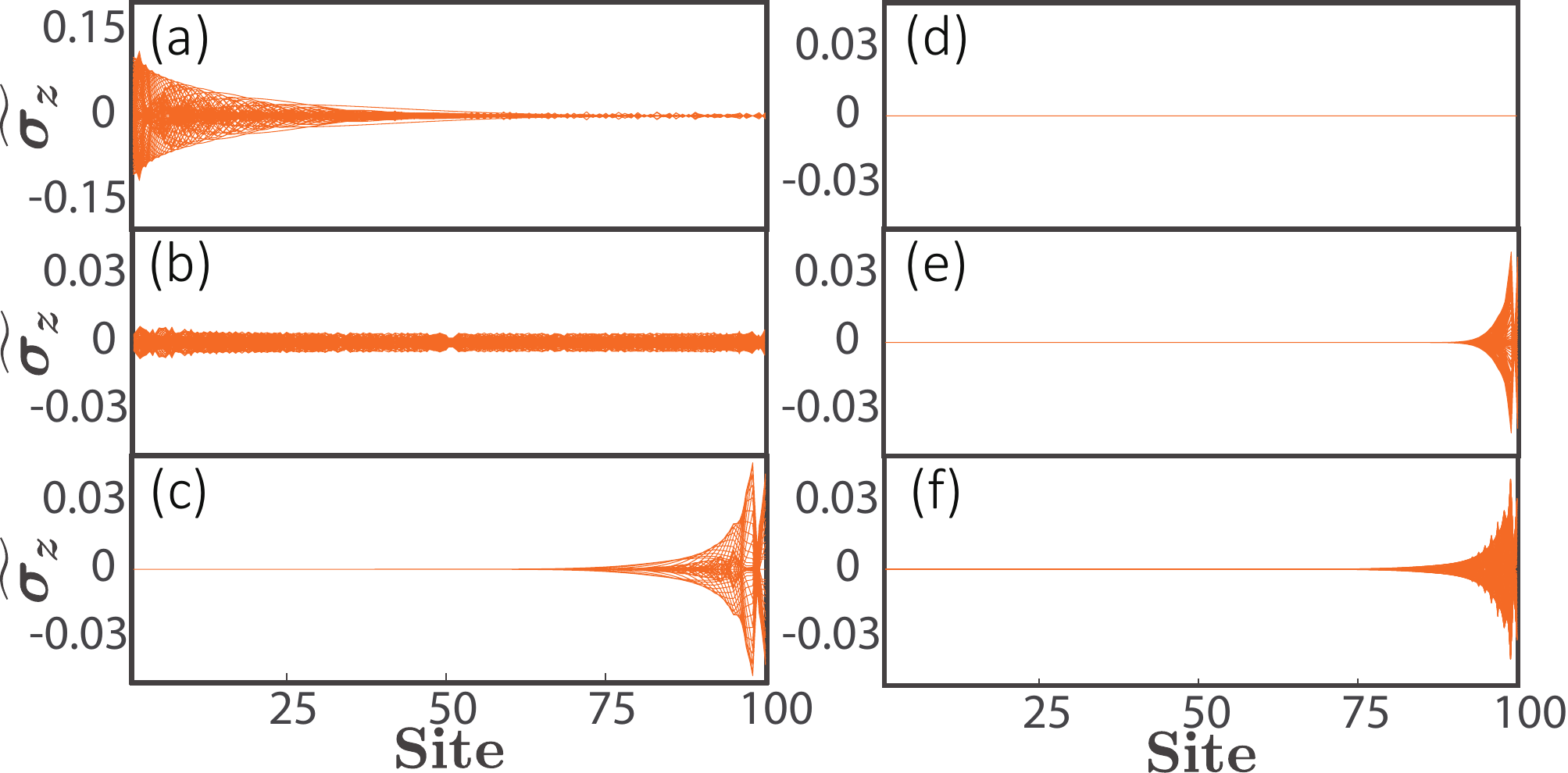}
\caption{(a) to (d) Evolution of the spin component $\tilde{\sigma}_z$ with increasing magnetic field amplitude along the dashed line in Fig. 3(b). The system parameters are $\theta=\frac{\pi}{3}$, $\phi=\frac{\pi}{4}$, and $\delta_x=$1, 1.57, 2, and 7 for (a) to (d), respectively. (e) and (f) The spin components $\tilde{\sigma}_z$ with $\delta_x=7$, $\theta=\frac{\pi}{2}$, and $\phi=\frac{\pi}{4}$, and $\delta_x=7$, $\theta=\frac{2\pi}{3}$, and $\phi=\frac{\pi}{4}$ in the transition point, and anti-$\mathcal{PT}$ symmetric phase, respectively.}
\label{fig5}
\end{figure}
In Figs. \ref{fig3}(c)-\ref{fig3}(f), we essentially show a non-Bloch $\mathcal{PT}$ transition within the spinful model of Eq. (\ref{modelh}). 
With increasing $\delta_x$, 
the OBC spectra change from partially to entirely real. 
Meanwhile, the corresponding spin polarizations of 
$\tilde{\sigma}_z$ Figs. \ref{fig5}(a)-\ref{fig5}(d). $\tilde{\sigma}_z=0$ indeed corresponds to the system under the exact non-Bloch $\mathcal{PT}$ phase [Figs. \ref{fig3}(f) and \ref{fig5}(d)]. In contrast, regardless of whether the system is under the phase transition point or in the non-Bloch anti-$\mathcal{PT}$ symmetric phase, $\tilde{\sigma}_z$ remains nonzero [Figs. \ref{fig5}(e) and \ref{fig5}(f)]. In general, the emergence of $\widetilde{\sigma }_z=0$, which is equivalent to all-real eigenenergies, is an origin of the non-Bloch $\mathcal{PT}$ symmetry in a general spinful system (SM Sec. VI. D \cite{sm}).

\textit{Conclusion.}\textemdash In this work, we showed that by introducing appropriate SU(2) phases, and applying an external magnetic field in a spinful non-Hermitian system, states can be skinned at both sides or one side 
and that magnetic fields can induce FOSE in non-Hermitian spinful systems. Such an anomalous phenomenon stems from the emergence of nonreciprocal hopping strengths induced by external magnetic fields, coupled to internal geometric asymmetries in the unit cell of our lattice. For sufficient magnetic field strength, we also demonstrated a transition between non-Bloch $\mathcal{PT}$ and anti-$\mathcal{PT}$ symmetric phases via the hopping phases under a strong external magnetic field. We used perturbation theory to derive the transition condition and elucidated the symmetry of the spin polarization in this system. In addition, we unveiled a unique signature of the non-Bloch $\mathcal{PT}$ symmetry, which provides a new way to understand the spin-dependent origin of non-Bloch $\mathcal{PT}$ symmetry in spinful non-Hermitian systems. Note that the model can be generalized to two dimensions, allowing us to control the localization of the states to any corner (see SM Sec. VII \cite{sm}).
Our results may be verified in physical platforms such as ultracold atoms and photonic systems, such as optical fibers. Artificial non-Abelian gauge potentials have been created to act on cold atoms in optical lattices \cite{Wang2012spin}. Based on this platform, the SU(2) gauge potentials introduced here can be synthesized \cite{yang2019synthesis,guo2021experimental,sun2022non,zhang2022non,cheng2023artificial,pang2024synthetic}, and highly tunable Zeeman couplings in cold atoms can be achieved through Raman lasers \cite{Xu2016dirac, lin2011spin}. Recent experiments have also realized spin-dependent hopping using a synthetic frequency dimension platform \cite{Cheng2024Non}. Remarkably, spin-related detection is a potential tool to observe the features of non-Hermitian systems, especially the skin effect with 
spin separation explored here. 

W.L. was supported by the NSF-China under Grant No. 11804396, Y.H. was supported by the Natural Science Foundation of Hunan Province Grant No. 2024JJ5426m, and A.A. was supported by the Simons Foundation. This work was supported in part by the High Performance Computing Center of Central South University.
	
\section*{Reference}
\makeatletter
\renewcommand{\bibsection}{}
\makeatother
\bibliographystyle{apsrev4-2}
\bibliography{ref}

\begin{thebibliography}{55}%
\makeatletter
\providecommand \@ifxundefined [1]{%
 \@ifx{#1\undefined}
}%
\providecommand \@ifnum [1]{%
 \ifnum #1\expandafter \@firstoftwo
 \else \expandafter \@secondoftwo
 \fi
}%
\providecommand \@ifx [1]{%
 \ifx #1\expandafter \@firstoftwo
 \else \expandafter \@secondoftwo
 \fi
}%
\providecommand \natexlab [1]{#1}%
\providecommand \enquote  [1]{``#1''}%
\providecommand \bibnamefont  [1]{#1}%
\providecommand \bibfnamefont [1]{#1}%
\providecommand \citenamefont [1]{#1}%
\providecommand \href@noop [0]{\@secondoftwo}%
\providecommand \href [0]{\begingroup \@sanitize@url \@href}%
\providecommand \@href[1]{\@@startlink{#1}\@@href}%
\providecommand \@@href[1]{\endgroup#1\@@endlink}%
\providecommand \@sanitize@url [0]{\catcode `\\12\catcode `\$12\catcode
  `\&12\catcode `\#12\catcode `\^12\catcode `\_12\catcode `\%12\relax}%
\providecommand \@@startlink[1]{}%
\providecommand \@@endlink[0]{}%
\providecommand \url  [0]{\begingroup\@sanitize@url \@url }%
\providecommand \@url [1]{\endgroup\@href {#1}{\urlprefix }}%
\providecommand \urlprefix  [0]{URL }%
\providecommand \Eprint [0]{\href }%
\providecommand \doibase [0]{https://doi.org/}%
\providecommand \selectlanguage [0]{\@gobble}%
\providecommand \bibinfo  [0]{\@secondoftwo}%
\providecommand \bibfield  [0]{\@secondoftwo}%
\providecommand \translation [1]{[#1]}%
\providecommand \BibitemOpen [0]{}%
\providecommand \bibitemStop [0]{}%
\providecommand \bibitemNoStop [0]{.\EOS\space}%
\providecommand \EOS [0]{\spacefactor3000\relax}%
\providecommand \BibitemShut  [1]{\csname bibitem#1\endcsname}%
\let\auto@bib@innerbib\@empty
%</preamble>
\bibitem [{\citenamefont {Peng}\ \emph {et~al.}(2014)\citenamefont {Peng},
  \citenamefont {{\"O}zdemir}, \citenamefont {Lei}, \citenamefont {Monifi},
  \citenamefont {Gianfreda}, \citenamefont {Long}, \citenamefont {Fan},
  \citenamefont {Nori}, \citenamefont {Bender},\ and\ \citenamefont
  {Yang}}]{peng2014parity}%
  \BibitemOpen
  \bibfield  {author} {\bibinfo {author} {\bibfnamefont {B.}~\bibnamefont
  {Peng}}, \bibinfo {author} {\bibfnamefont {{\c{S}}.~K.}\ \bibnamefont
  {{\"O}zdemir}}, \bibinfo {author} {\bibfnamefont {F.}~\bibnamefont {Lei}},
  \bibinfo {author} {\bibfnamefont {F.}~\bibnamefont {Monifi}}, \bibinfo
  {author} {\bibfnamefont {M.}~\bibnamefont {Gianfreda}}, \bibinfo {author}
  {\bibfnamefont {G.~L.}\ \bibnamefont {Long}}, \bibinfo {author}
  {\bibfnamefont {S.}~\bibnamefont {Fan}}, \bibinfo {author} {\bibfnamefont
  {F.}~\bibnamefont {Nori}}, \bibinfo {author} {\bibfnamefont {C.~M.}\
  \bibnamefont {Bender}},\ and\ \bibinfo {author} {\bibfnamefont
  {L.}~\bibnamefont {Yang}},\ }\href {https://doi.org/10.1038/nphys2927}
  {\bibfield  {journal} {\bibinfo  {journal} {Nat. Phys.}\ }\textbf {\bibinfo
  {volume} {10}},\ \bibinfo {pages} {394} (\bibinfo {year} {2014})}\BibitemShut
  {NoStop}%
\bibitem [{\citenamefont {Ramezani}\ \emph {et~al.}(2010)\citenamefont
  {Ramezani}, \citenamefont {Kottos}, \citenamefont {El-Ganainy},\ and\
  \citenamefont {Christodoulides}}]{ramezani2010unidirectional}%
  \BibitemOpen
  \bibfield  {author} {\bibinfo {author} {\bibfnamefont {H.}~\bibnamefont
  {Ramezani}}, \bibinfo {author} {\bibfnamefont {T.}~\bibnamefont {Kottos}},
  \bibinfo {author} {\bibfnamefont {R.}~\bibnamefont {El-Ganainy}},\ and\
  \bibinfo {author} {\bibfnamefont {D.~N.}\ \bibnamefont {Christodoulides}},\
  }\href {https://doi.org/10.1103/PhysRevA.82.043803} {\bibfield  {journal}
  {\bibinfo  {journal} {Phys. Rev. A}\ }\textbf {\bibinfo {volume} {82}},\
  \bibinfo {pages} {043803} (\bibinfo {year} {2010})}\BibitemShut {NoStop}%
\bibitem [{\citenamefont {Chang}\ \emph {et~al.}(2014)\citenamefont {Chang},
  \citenamefont {Jiang}, \citenamefont {Hua}, \citenamefont {Yang},
  \citenamefont {Wen}, \citenamefont {Jiang}, \citenamefont {Li}, \citenamefont
  {Wang},\ and\ \citenamefont {Xiao}}]{chang2014parity}%
  \BibitemOpen
  \bibfield  {author} {\bibinfo {author} {\bibfnamefont {L.}~\bibnamefont
  {Chang}}, \bibinfo {author} {\bibfnamefont {X.}~\bibnamefont {Jiang}},
  \bibinfo {author} {\bibfnamefont {S.}~\bibnamefont {Hua}}, \bibinfo {author}
  {\bibfnamefont {C.}~\bibnamefont {Yang}}, \bibinfo {author} {\bibfnamefont
  {J.}~\bibnamefont {Wen}}, \bibinfo {author} {\bibfnamefont {L.}~\bibnamefont
  {Jiang}}, \bibinfo {author} {\bibfnamefont {G.}~\bibnamefont {Li}}, \bibinfo
  {author} {\bibfnamefont {G.}~\bibnamefont {Wang}},\ and\ \bibinfo {author}
  {\bibfnamefont {M.}~\bibnamefont {Xiao}},\ }\href
  {https://doi.org/10.1038/nphoton.2014.133} {\bibfield  {journal} {\bibinfo
  {journal} {Nat. Photonics}\ }\textbf {\bibinfo {volume} {8}},\ \bibinfo
  {pages} {524} (\bibinfo {year} {2014})}\BibitemShut {NoStop}%
\bibitem [{\citenamefont {Chen}\ \emph {et~al.}(2016)\citenamefont {Chen},
  \citenamefont {Liu}, \citenamefont {Evans}, \citenamefont {Paulose},
  \citenamefont {Cohen}, \citenamefont {Vitelli},\ and\ \citenamefont
  {Santangelo}}]{chen2016topological}%
  \BibitemOpen
  \bibfield  {author} {\bibinfo {author} {\bibfnamefont {B.~G.}\ \bibnamefont
  {Chen}}, \bibinfo {author} {\bibfnamefont {B.}~\bibnamefont {Liu}}, \bibinfo
  {author} {\bibfnamefont {A.~A.}\ \bibnamefont {Evans}}, \bibinfo {author}
  {\bibfnamefont {J.}~\bibnamefont {Paulose}}, \bibinfo {author} {\bibfnamefont
  {I.}~\bibnamefont {Cohen}}, \bibinfo {author} {\bibfnamefont
  {V.}~\bibnamefont {Vitelli}},\ and\ \bibinfo {author} {\bibfnamefont {C.~D.}\
  \bibnamefont {Santangelo}},\ }\href
  {https://doi.org/10.1103/PhysRevLett.116.135501} {\bibfield  {journal}
  {\bibinfo  {journal} {Phys. Rev. Lett.}\ }\textbf {\bibinfo {volume} {116}},\
  \bibinfo {pages} {135501} (\bibinfo {year} {2016})}\BibitemShut {NoStop}%
\bibitem [{\citenamefont {Zhou}\ \emph {et~al.}(2018)\citenamefont {Zhou},
  \citenamefont {Peng}, \citenamefont {Yoon}, \citenamefont {Hsu},
  \citenamefont {Nelson}, \citenamefont {Fu}, \citenamefont {Joannopoulos},
  \citenamefont {Solja{\v{c}}i{\'c}},\ and\ \citenamefont
  {Zhen}}]{zhou2018observation}%
  \BibitemOpen
  \bibfield  {author} {\bibinfo {author} {\bibfnamefont {H.}~\bibnamefont
  {Zhou}}, \bibinfo {author} {\bibfnamefont {C.}~\bibnamefont {Peng}}, \bibinfo
  {author} {\bibfnamefont {Y.}~\bibnamefont {Yoon}}, \bibinfo {author}
  {\bibfnamefont {C.~W.}\ \bibnamefont {Hsu}}, \bibinfo {author} {\bibfnamefont
  {K.~A.}\ \bibnamefont {Nelson}}, \bibinfo {author} {\bibfnamefont
  {L.}~\bibnamefont {Fu}}, \bibinfo {author} {\bibfnamefont {J.~D.}\
  \bibnamefont {Joannopoulos}}, \bibinfo {author} {\bibfnamefont
  {M.}~\bibnamefont {Solja{\v{c}}i{\'c}}},\ and\ \bibinfo {author}
  {\bibfnamefont {B.}~\bibnamefont {Zhen}},\ }\href
  {https://doi.org/10.1126/science.aap9859} {\bibfield  {journal} {\bibinfo
  {journal} {Science}\ }\textbf {\bibinfo {volume} {359}},\ \bibinfo {pages}
  {1009} (\bibinfo {year} {2018})}\BibitemShut {NoStop}%
\bibitem [{\citenamefont {Shen}\ and\ \citenamefont
  {Fu}(2018)}]{shen2018quantum}%
  \BibitemOpen
  \bibfield  {author} {\bibinfo {author} {\bibfnamefont {H.}~\bibnamefont
  {Shen}}\ and\ \bibinfo {author} {\bibfnamefont {L.}~\bibnamefont {Fu}},\
  }\href {https://doi.org/10.1103/PhysRevLett.121.026403} {\bibfield  {journal}
  {\bibinfo  {journal} {Phys. Rev. Lett.}\ }\textbf {\bibinfo {volume} {121}},\
  \bibinfo {pages} {026403} (\bibinfo {year} {2018})}\BibitemShut {NoStop}%
\bibitem [{\citenamefont {Papaj}\ \emph {et~al.}(2019)\citenamefont {Papaj},
  \citenamefont {Isobe},\ and\ \citenamefont {Fu}}]{papaj2019nodal}%
  \BibitemOpen
  \bibfield  {author} {\bibinfo {author} {\bibfnamefont {M.}~\bibnamefont
  {Papaj}}, \bibinfo {author} {\bibfnamefont {H.}~\bibnamefont {Isobe}},\ and\
  \bibinfo {author} {\bibfnamefont {L.}~\bibnamefont {Fu}},\ }\href
  {https://doi.org/10.1103/PhysRevB.99.201107} {\bibfield  {journal} {\bibinfo
  {journal} {Phys. Rev. B}\ }\textbf {\bibinfo {volume} {99}},\ \bibinfo
  {pages} {201107} (\bibinfo {year} {2019})}\BibitemShut {NoStop}%
\bibitem [{\citenamefont {Yoshida}\ \emph {et~al.}(2018)\citenamefont
  {Yoshida}, \citenamefont {Peters},\ and\ \citenamefont
  {Kawakami}}]{yoshida2018non}%
  \BibitemOpen
  \bibfield  {author} {\bibinfo {author} {\bibfnamefont {T.}~\bibnamefont
  {Yoshida}}, \bibinfo {author} {\bibfnamefont {R.}~\bibnamefont {Peters}},\
  and\ \bibinfo {author} {\bibfnamefont {N.}~\bibnamefont {Kawakami}},\ }\href
  {https://doi.org/10.1103/PhysRevB.98.035141} {\bibfield  {journal} {\bibinfo
  {journal} {Phys. Rev. B}\ }\textbf {\bibinfo {volume} {98}},\ \bibinfo
  {pages} {035141} (\bibinfo {year} {2018})}\BibitemShut {NoStop}%
\bibitem [{\citenamefont {Yao}\ and\ \citenamefont {Wang}(2018)}]{yao2018edge}%
  \BibitemOpen
  \bibfield  {author} {\bibinfo {author} {\bibfnamefont {S.}~\bibnamefont
  {Yao}}\ and\ \bibinfo {author} {\bibfnamefont {Z.}~\bibnamefont {Wang}},\
  }\href {https://doi.org/10.1103/PhysRevLett.121.086803} {\bibfield  {journal}
  {\bibinfo  {journal} {Phys. Rev. Lett.}\ }\textbf {\bibinfo {volume} {121}},\
  \bibinfo {pages} {086803} (\bibinfo {year} {2018})}\BibitemShut {NoStop}%
\bibitem [{\citenamefont {Kawabata}\ \emph {et~al.}(2019)\citenamefont
  {Kawabata}, \citenamefont {Shiozaki}, \citenamefont {Ueda},\ and\
  \citenamefont {Sato}}]{PhysRevX.9.041015}%
  \BibitemOpen
  \bibfield  {author} {\bibinfo {author} {\bibfnamefont {K.}~\bibnamefont
  {Kawabata}}, \bibinfo {author} {\bibfnamefont {K.}~\bibnamefont {Shiozaki}},
  \bibinfo {author} {\bibfnamefont {M.}~\bibnamefont {Ueda}},\ and\ \bibinfo
  {author} {\bibfnamefont {M.}~\bibnamefont {Sato}},\ }\href
  {https://doi.org/10.1103/PhysRevX.9.041015} {\bibfield  {journal} {\bibinfo
  {journal} {Phys. Rev. X}\ }\textbf {\bibinfo {volume} {9}},\ \bibinfo {pages}
  {041015} (\bibinfo {year} {2019})}\BibitemShut {NoStop}%
\bibitem [{\citenamefont {Zhang}\ \emph
  {et~al.}(2022{\natexlab{a}})\citenamefont {Zhang}, \citenamefont {Zhang},
  \citenamefont {Lu},\ and\ \citenamefont {Chen}}]{zhang2022review}%
  \BibitemOpen
  \bibfield  {author} {\bibinfo {author} {\bibfnamefont {X.}~\bibnamefont
  {Zhang}}, \bibinfo {author} {\bibfnamefont {T.}~\bibnamefont {Zhang}},
  \bibinfo {author} {\bibfnamefont {M.~H.}\ \bibnamefont {Lu}},\ and\ \bibinfo
  {author} {\bibfnamefont {Y.~F.}\ \bibnamefont {Chen}},\ }\href
  {https://doi.org/10.1080/23746149.2022.2109431} {\bibfield  {journal}
  {\bibinfo  {journal} {Advances in Physics: X}\ }\textbf {\bibinfo {volume}
  {7}},\ \bibinfo {pages} {2109431} (\bibinfo {year}
  {2022}{\natexlab{a}})}\BibitemShut {NoStop}%
\bibitem [{\citenamefont {Borgnia}\ \emph {et~al.}(2020)\citenamefont
  {Borgnia}, \citenamefont {Kruchkov},\ and\ \citenamefont {Slager}}]{slager}%
  \BibitemOpen
  \bibfield  {author} {\bibinfo {author} {\bibfnamefont {D.~S.}\ \bibnamefont
  {Borgnia}}, \bibinfo {author} {\bibfnamefont {A.~J.}\ \bibnamefont
  {Kruchkov}},\ and\ \bibinfo {author} {\bibfnamefont {R.-J.}\ \bibnamefont
  {Slager}},\ }\href {https://doi.org/10.1103/PhysRevLett.124.056802}
  {\bibfield  {journal} {\bibinfo  {journal} {Phys. Rev. Lett.}\ }\textbf
  {\bibinfo {volume} {124}},\ \bibinfo {pages} {056802} (\bibinfo {year}
  {2020})}\BibitemShut {NoStop}%
\bibitem [{\citenamefont {Zhang}\ \emph {et~al.}(2020)\citenamefont {Zhang},
  \citenamefont {Yang},\ and\ \citenamefont {Fang}}]{Zhang2021Correspondence}%
  \BibitemOpen
  \bibfield  {author} {\bibinfo {author} {\bibfnamefont {K.}~\bibnamefont
  {Zhang}}, \bibinfo {author} {\bibfnamefont {Z.}~\bibnamefont {Yang}},\ and\
  \bibinfo {author} {\bibfnamefont {C.}~\bibnamefont {Fang}},\ }\href
  {https://doi.org/10.1103/PhysRevLett.125.126402} {\bibfield  {journal}
  {\bibinfo  {journal} {Phys. Rev. Lett.}\ }\textbf {\bibinfo {volume} {125}},\
  \bibinfo {pages} {126402} (\bibinfo {year} {2020})}\BibitemShut {NoStop}%
\bibitem [{\citenamefont {Kunst}\ \emph {et~al.}(2018)\citenamefont {Kunst},
  \citenamefont {Edvardsson}, \citenamefont {Budich},\ and\ \citenamefont
  {Bergholtz}}]{kunst2018biorthogonal}%
  \BibitemOpen
  \bibfield  {author} {\bibinfo {author} {\bibfnamefont {F.~K.}\ \bibnamefont
  {Kunst}}, \bibinfo {author} {\bibfnamefont {E.}~\bibnamefont {Edvardsson}},
  \bibinfo {author} {\bibfnamefont {J.~C.}\ \bibnamefont {Budich}},\ and\
  \bibinfo {author} {\bibfnamefont {E.~J.}\ \bibnamefont {Bergholtz}},\ }\href
  {https://doi.org/10.1103/PhysRevLett.121.026808} {\bibfield  {journal}
  {\bibinfo  {journal} {Phys. Rev. Lett.}\ }\textbf {\bibinfo {volume} {121}},\
  \bibinfo {pages} {026808} (\bibinfo {year} {2018})}\BibitemShut {NoStop}%
\bibitem [{\citenamefont {Li}\ \emph {et~al.}(2022)\citenamefont {Li},
  \citenamefont {Liang}, \citenamefont {Wang}, \citenamefont {Lu},\ and\
  \citenamefont {Liu}}]{Li2022Gain}%
  \BibitemOpen
  \bibfield  {author} {\bibinfo {author} {\bibfnamefont {Y.}~\bibnamefont
  {Li}}, \bibinfo {author} {\bibfnamefont {C.}~\bibnamefont {Liang}}, \bibinfo
  {author} {\bibfnamefont {C.}~\bibnamefont {Wang}}, \bibinfo {author}
  {\bibfnamefont {C.}~\bibnamefont {Lu}},\ and\ \bibinfo {author}
  {\bibfnamefont {Y.~C.}\ \bibnamefont {Liu}},\ }\href
  {https://doi.org/10.1103/PhysRevLett.128.223903} {\bibfield  {journal}
  {\bibinfo  {journal} {Phys. Rev. Lett.}\ }\textbf {\bibinfo {volume} {128}},\
  \bibinfo {pages} {223903} (\bibinfo {year} {2022})}\BibitemShut {NoStop}%
\bibitem [{\citenamefont {Weidemann}\ \emph {et~al.}(2020)\citenamefont
  {Weidemann}, \citenamefont {Kremer}, \citenamefont {Helbig}, \citenamefont
  {Hofmann}, \citenamefont {Stegmaier}, \citenamefont {Greiter}, \citenamefont
  {Thomale},\ and\ \citenamefont {Szameit}}]{weidemann2020topological}%
  \BibitemOpen
  \bibfield  {author} {\bibinfo {author} {\bibfnamefont {S.}~\bibnamefont
  {Weidemann}}, \bibinfo {author} {\bibfnamefont {M.}~\bibnamefont {Kremer}},
  \bibinfo {author} {\bibfnamefont {T.}~\bibnamefont {Helbig}}, \bibinfo
  {author} {\bibfnamefont {T.}~\bibnamefont {Hofmann}}, \bibinfo {author}
  {\bibfnamefont {A.}~\bibnamefont {Stegmaier}}, \bibinfo {author}
  {\bibfnamefont {M.}~\bibnamefont {Greiter}}, \bibinfo {author} {\bibfnamefont
  {R.}~\bibnamefont {Thomale}},\ and\ \bibinfo {author} {\bibfnamefont
  {A.}~\bibnamefont {Szameit}},\ }\href
  {https://doi.org/10.1126/science.aaz8727} {\bibfield  {journal} {\bibinfo
  {journal} {Science}\ }\textbf {\bibinfo {volume} {368}},\ \bibinfo {pages}
  {311} (\bibinfo {year} {2020})}\BibitemShut {NoStop}%
\bibitem [{\citenamefont {Wang}\ \emph {et~al.}(2021)\citenamefont {Wang},
  \citenamefont {Dutt}, \citenamefont {Yang}, \citenamefont {Wojcik},
  \citenamefont {Vu\v{c}kovi\'c},\ and\ \citenamefont
  {Fan}}]{wang2021generating}%
  \BibitemOpen
  \bibfield  {author} {\bibinfo {author} {\bibfnamefont {K.}~\bibnamefont
  {Wang}}, \bibinfo {author} {\bibfnamefont {A.}~\bibnamefont {Dutt}}, \bibinfo
  {author} {\bibfnamefont {K.~Y.}\ \bibnamefont {Yang}}, \bibinfo {author}
  {\bibfnamefont {C.~C.}\ \bibnamefont {Wojcik}}, \bibinfo {author}
  {\bibfnamefont {J.}~\bibnamefont {Vu\v{c}kovi\'c}},\ and\ \bibinfo {author}
  {\bibfnamefont {S.}~\bibnamefont {Fan}},\ }\href
  {https://doi.org/10.1126/science.abf6568} {\bibfield  {journal} {\bibinfo
  {journal} {Science}\ }\textbf {\bibinfo {volume} {371}},\ \bibinfo {pages}
  {1240} (\bibinfo {year} {2021})}\BibitemShut {NoStop}%
\bibitem [{\citenamefont {Xiao}\ \emph {et~al.}(2020)\citenamefont {Xiao},
  \citenamefont {Deng}, \citenamefont {Wang}, \citenamefont {Zhu},
  \citenamefont {Wang}, \citenamefont {Yi},\ and\ \citenamefont
  {Xue}}]{xiao2020non}%
  \BibitemOpen
  \bibfield  {author} {\bibinfo {author} {\bibfnamefont {L.}~\bibnamefont
  {Xiao}}, \bibinfo {author} {\bibfnamefont {T.}~\bibnamefont {Deng}}, \bibinfo
  {author} {\bibfnamefont {K.}~\bibnamefont {Wang}}, \bibinfo {author}
  {\bibfnamefont {G.}~\bibnamefont {Zhu}}, \bibinfo {author} {\bibfnamefont
  {Z.}~\bibnamefont {Wang}}, \bibinfo {author} {\bibfnamefont {W.}~\bibnamefont
  {Yi}},\ and\ \bibinfo {author} {\bibfnamefont {P.}~\bibnamefont {Xue}},\
  }\href {https://doi.org/10.1038/s41567-020-0836-6} {\bibfield  {journal}
  {\bibinfo  {journal} {Nat. Phys.}\ }\textbf {\bibinfo {volume} {16}},\
  \bibinfo {pages} {761} (\bibinfo {year} {2020})}\BibitemShut {NoStop}%
\bibitem [{\citenamefont {Zhang}\ \emph
  {et~al.}(2021{\natexlab{a}})\citenamefont {Zhang}, \citenamefont {Tian},
  \citenamefont {Jiang}, \citenamefont {Lu},\ and\ \citenamefont
  {Chen}}]{zhang2021observation}%
  \BibitemOpen
  \bibfield  {author} {\bibinfo {author} {\bibfnamefont {X.}~\bibnamefont
  {Zhang}}, \bibinfo {author} {\bibfnamefont {Y.}~\bibnamefont {Tian}},
  \bibinfo {author} {\bibfnamefont {J.~H.}\ \bibnamefont {Jiang}}, \bibinfo
  {author} {\bibfnamefont {M.~H.}\ \bibnamefont {Lu}},\ and\ \bibinfo {author}
  {\bibfnamefont {Y.~F.}\ \bibnamefont {Chen}},\ }\href
  {https://doi.org/10.1038/s41467-021-25716-y} {\bibfield  {journal} {\bibinfo
  {journal} {Nat. Commun.}\ }\textbf {\bibinfo {volume} {12}},\ \bibinfo
  {pages} {5377} (\bibinfo {year} {2021}{\natexlab{a}})}\BibitemShut {NoStop}%
\bibitem [{\citenamefont {Zhang}\ \emph
  {et~al.}(2021{\natexlab{b}})\citenamefont {Zhang}, \citenamefont {Yang},
  \citenamefont {Ge}, \citenamefont {Guan}, \citenamefont {Chen}, \citenamefont
  {Yan}, \citenamefont {Chen}, \citenamefont {Xi}, \citenamefont {Li},
  \citenamefont {Jia} \emph {et~al.}}]{zhang2021acoustic}%
  \BibitemOpen
  \bibfield  {author} {\bibinfo {author} {\bibfnamefont {L.}~\bibnamefont
  {Zhang}}, \bibinfo {author} {\bibfnamefont {Y.}~\bibnamefont {Yang}},
  \bibinfo {author} {\bibfnamefont {Y.}~\bibnamefont {Ge}}, \bibinfo {author}
  {\bibfnamefont {Y.~J.}\ \bibnamefont {Guan}}, \bibinfo {author}
  {\bibfnamefont {Q.}~\bibnamefont {Chen}}, \bibinfo {author} {\bibfnamefont
  {Q.}~\bibnamefont {Yan}}, \bibinfo {author} {\bibfnamefont {F.}~\bibnamefont
  {Chen}}, \bibinfo {author} {\bibfnamefont {R.}~\bibnamefont {Xi}}, \bibinfo
  {author} {\bibfnamefont {Y.}~\bibnamefont {Li}}, \bibinfo {author}
  {\bibfnamefont {D.}~\bibnamefont {Jia}}, \emph {et~al.},\ }\href
  {https://doi.org/10.1038/s41467-021-26619-8} {\bibfield  {journal} {\bibinfo
  {journal} {Nat. Commun.}\ }\textbf {\bibinfo {volume} {12}},\ \bibinfo
  {pages} {6297} (\bibinfo {year} {2021}{\natexlab{b}})}\BibitemShut {NoStop}%
\bibitem [{\citenamefont {Gu}\ \emph {et~al.}(2022)\citenamefont {Gu},
  \citenamefont {Gao}, \citenamefont {Xue}, \citenamefont {Li}, \citenamefont
  {Su},\ and\ \citenamefont {Zhu}}]{gu2022transient}%
  \BibitemOpen
  \bibfield  {author} {\bibinfo {author} {\bibfnamefont {Z.}~\bibnamefont
  {Gu}}, \bibinfo {author} {\bibfnamefont {H.}~\bibnamefont {Gao}}, \bibinfo
  {author} {\bibfnamefont {H.}~\bibnamefont {Xue}}, \bibinfo {author}
  {\bibfnamefont {J.}~\bibnamefont {Li}}, \bibinfo {author} {\bibfnamefont
  {Z.}~\bibnamefont {Su}},\ and\ \bibinfo {author} {\bibfnamefont
  {J.}~\bibnamefont {Zhu}},\ }\href
  {https://doi.org/10.1038/s41467-022-35448-2} {\bibfield  {journal} {\bibinfo
  {journal} {Nat. Commun.}\ }\textbf {\bibinfo {volume} {13}},\ \bibinfo
  {pages} {7668} (\bibinfo {year} {2022})}\BibitemShut {NoStop}%
\bibitem [{\citenamefont {Wang}\ \emph {et~al.}(2022)\citenamefont {Wang},
  \citenamefont {Wang},\ and\ \citenamefont {Ma}}]{wang2022non}%
  \BibitemOpen
  \bibfield  {author} {\bibinfo {author} {\bibfnamefont {W.}~\bibnamefont
  {Wang}}, \bibinfo {author} {\bibfnamefont {X.}~\bibnamefont {Wang}},\ and\
  \bibinfo {author} {\bibfnamefont {G.}~\bibnamefont {Ma}},\ }\href
  {https://doi.org/10.1038/s41586-022-04929-1} {\bibfield  {journal} {\bibinfo
  {journal} {Nature}\ }\textbf {\bibinfo {volume} {608}},\ \bibinfo {pages}
  {50} (\bibinfo {year} {2022})}\BibitemShut {NoStop}%
\bibitem [{\citenamefont {Li}\ \emph {et~al.}(2020)\citenamefont {Li},
  \citenamefont {Lee}, \citenamefont {Mu},\ and\ \citenamefont
  {Gong}}]{li2020critical}%
  \BibitemOpen
  \bibfield  {author} {\bibinfo {author} {\bibfnamefont {L.}~\bibnamefont
  {Li}}, \bibinfo {author} {\bibfnamefont {C.~H.}\ \bibnamefont {Lee}},
  \bibinfo {author} {\bibfnamefont {S.}~\bibnamefont {Mu}},\ and\ \bibinfo
  {author} {\bibfnamefont {J.}~\bibnamefont {Gong}},\ }\href
  {https://doi.org/10.1038/s41467-020-18917-4} {\bibfield  {journal} {\bibinfo
  {journal} {Nat. Commun.}\ }\textbf {\bibinfo {volume} {11}},\ \bibinfo
  {pages} {5491} (\bibinfo {year} {2020})}\BibitemShut {NoStop}%
\bibitem [{\citenamefont {Helbig}\ \emph {et~al.}(2020)\citenamefont {Helbig},
  \citenamefont {Hofmann}, \citenamefont {Imhof}, \citenamefont {Abdelghany},
  \citenamefont {Kiessling}, \citenamefont {Molenkamp}, \citenamefont {Lee},
  \citenamefont {Szameit}, \citenamefont {Greiter},\ and\ \citenamefont
  {Thomale}}]{helbig2020generalized}%
  \BibitemOpen
  \bibfield  {author} {\bibinfo {author} {\bibfnamefont {T.}~\bibnamefont
  {Helbig}}, \bibinfo {author} {\bibfnamefont {T.}~\bibnamefont {Hofmann}},
  \bibinfo {author} {\bibfnamefont {S.}~\bibnamefont {Imhof}}, \bibinfo
  {author} {\bibfnamefont {M.}~\bibnamefont {Abdelghany}}, \bibinfo {author}
  {\bibfnamefont {T.}~\bibnamefont {Kiessling}}, \bibinfo {author}
  {\bibfnamefont {L.}~\bibnamefont {Molenkamp}}, \bibinfo {author}
  {\bibfnamefont {C.}~\bibnamefont {Lee}}, \bibinfo {author} {\bibfnamefont
  {A.}~\bibnamefont {Szameit}}, \bibinfo {author} {\bibfnamefont
  {M.}~\bibnamefont {Greiter}},\ and\ \bibinfo {author} {\bibfnamefont
  {R.}~\bibnamefont {Thomale}},\ }\href
  {https://doi.org/10.1038/s41567-020-0922-9} {\bibfield  {journal} {\bibinfo
  {journal} {Nat. Phys.}\ }\textbf {\bibinfo {volume} {16}},\ \bibinfo {pages}
  {747} (\bibinfo {year} {2020})}\BibitemShut {NoStop}%
\bibitem [{\citenamefont {Zhu}\ \emph {et~al.}(2022)\citenamefont {Zhu},
  \citenamefont {Wang}, \citenamefont {Leykam}, \citenamefont {Xue},
  \citenamefont {Wang},\ and\ \citenamefont {Chong}}]{zhu2022anomalous}%
  \BibitemOpen
  \bibfield  {author} {\bibinfo {author} {\bibfnamefont {B.}~\bibnamefont
  {Zhu}}, \bibinfo {author} {\bibfnamefont {Q.}~\bibnamefont {Wang}}, \bibinfo
  {author} {\bibfnamefont {D.}~\bibnamefont {Leykam}}, \bibinfo {author}
  {\bibfnamefont {H.}~\bibnamefont {Xue}}, \bibinfo {author} {\bibfnamefont
  {Q.~J.}\ \bibnamefont {Wang}},\ and\ \bibinfo {author} {\bibfnamefont
  {Y.~D.}\ \bibnamefont {Chong}},\ }\href
  {https://doi.org/10.1103/PhysRevLett.129.013903} {\bibfield  {journal}
  {\bibinfo  {journal} {Phys. Rev. Lett.}\ }\textbf {\bibinfo {volume} {129}},\
  \bibinfo {pages} {013903} (\bibinfo {year} {2022})}\BibitemShut {NoStop}%
\bibitem [{\citenamefont {Franca}\ \emph {et~al.}(2022)\citenamefont {Franca},
  \citenamefont {Könye}, \citenamefont {Hassler}, \citenamefont {Brink},\ and\
  \citenamefont {Fulga}}]{franca2022non}%
  \BibitemOpen
  \bibfield  {author} {\bibinfo {author} {\bibfnamefont {S.}~\bibnamefont
  {Franca}}, \bibinfo {author} {\bibfnamefont {V.}~\bibnamefont {Könye}},
  \bibinfo {author} {\bibfnamefont {F.}~\bibnamefont {Hassler}}, \bibinfo
  {author} {\bibfnamefont {J.~v.~d.}\ \bibnamefont {Brink}},\ and\ \bibinfo
  {author} {\bibfnamefont {C.}~\bibnamefont {Fulga}},\ }\href
  {https://doi.org/10.1103/PhysRevLett.129.086601} {\bibfield  {journal}
  {\bibinfo  {journal} {Phys. Rev. Lett.}\ }\textbf {\bibinfo {volume} {129}},\
  \bibinfo {pages} {086601} (\bibinfo {year} {2022})}\BibitemShut {NoStop}%
\bibitem [{\citenamefont {Longhi}(2022)}]{Longhi2022Self}%
  \BibitemOpen
  \bibfield  {author} {\bibinfo {author} {\bibfnamefont {S.}~\bibnamefont
  {Longhi}},\ }\href {https://doi.org/10.1103/PhysRevLett.128.157601}
  {\bibfield  {journal} {\bibinfo  {journal} {Phys. Rev. Lett.}\ }\textbf
  {\bibinfo {volume} {128}},\ \bibinfo {pages} {157601} (\bibinfo {year}
  {2022})}\BibitemShut {NoStop}%
\bibitem [{\citenamefont {Lu}\ \emph {et~al.}(2021)\citenamefont {Lu},
  \citenamefont {Zhang},\ and\ \citenamefont {Franz}}]{lu2021magnetic}%
  \BibitemOpen
  \bibfield  {author} {\bibinfo {author} {\bibfnamefont {M.}~\bibnamefont
  {Lu}}, \bibinfo {author} {\bibfnamefont {X.-X.}\ \bibnamefont {Zhang}},\ and\
  \bibinfo {author} {\bibfnamefont {M.}~\bibnamefont {Franz}},\ }\href
  {https://doi.org/10.1103/PhysRevLett.127.256402} {\bibfield  {journal}
  {\bibinfo  {journal} {Phys. Rev. Lett.}\ }\textbf {\bibinfo {volume} {127}},\
  \bibinfo {pages} {256402} (\bibinfo {year} {2021})}\BibitemShut {NoStop}%
\bibitem [{\citenamefont {Shao}\ \emph {et~al.}(2022)\citenamefont {Shao},
  \citenamefont {Cai}, \citenamefont {Geng}, \citenamefont {Chen},\ and\
  \citenamefont {Xing}}]{Shao2022Cyclotron}%
  \BibitemOpen
  \bibfield  {author} {\bibinfo {author} {\bibfnamefont {K.}~\bibnamefont
  {Shao}}, \bibinfo {author} {\bibfnamefont {Z.~T.}\ \bibnamefont {Cai}},
  \bibinfo {author} {\bibfnamefont {H.}~\bibnamefont {Geng}}, \bibinfo {author}
  {\bibfnamefont {W.}~\bibnamefont {Chen}},\ and\ \bibinfo {author}
  {\bibfnamefont {D.~Y.}\ \bibnamefont {Xing}},\ }\href
  {https://doi.org/10.1103/PhysRevB.106.L081402} {\bibfield  {journal}
  {\bibinfo  {journal} {Phys. Rev. B}\ }\textbf {\bibinfo {volume} {106}},\
  \bibinfo {pages} {L081402} (\bibinfo {year} {2022})}\BibitemShut {NoStop}%
\bibitem [{\citenamefont {Li}\ \emph {et~al.}(2023)\citenamefont {Li},
  \citenamefont {Trauzettel}, \citenamefont {Neupert},\ and\ \citenamefont
  {Zhang}}]{Li2023enhancement}%
  \BibitemOpen
  \bibfield  {author} {\bibinfo {author} {\bibfnamefont {C.~A.}\ \bibnamefont
  {Li}}, \bibinfo {author} {\bibfnamefont {B.}~\bibnamefont {Trauzettel}},
  \bibinfo {author} {\bibfnamefont {T.}~\bibnamefont {Neupert}},\ and\ \bibinfo
  {author} {\bibfnamefont {S.~B.}\ \bibnamefont {Zhang}},\ }\href
  {https://doi.org/10.1103/PhysRevLett.131.116601} {\bibfield  {journal}
  {\bibinfo  {journal} {Phys. Rev. Lett.}\ }\textbf {\bibinfo {volume} {131}},\
  \bibinfo {pages} {116601} (\bibinfo {year} {2023})}\BibitemShut {NoStop}%
\bibitem [{\citenamefont {Asb{\'o}th}\ \emph {et~al.}(2016)\citenamefont
  {Asb{\'o}th}, \citenamefont {Oroszl{\'a}ny},\ and\ \citenamefont
  {P{\'a}lyi}}]{asboth2016short}%
  \BibitemOpen
  \bibfield  {author} {\bibinfo {author} {\bibfnamefont {J.~K.}\ \bibnamefont
  {Asb{\'o}th}}, \bibinfo {author} {\bibfnamefont {L.}~\bibnamefont
  {Oroszl{\'a}ny}},\ and\ \bibinfo {author} {\bibfnamefont {A.}~\bibnamefont
  {P{\'a}lyi}},\ }\href
  {https://link.springer.com/book/10.1007/978-3-319-25607-8} {\emph {\bibinfo
  {title} {A short course on topological insulators}}},\ Vol.\ \bibinfo
  {volume} {919}\ (\bibinfo  {publisher} {Springer},\ \bibinfo {year} {2016})\
  p.\ \bibinfo {pages} {166}\BibitemShut {NoStop}%
\bibitem [{\citenamefont {Hu}\ \emph {et~al.}(2024)\citenamefont {Hu},
  \citenamefont {Wang}, \citenamefont {Wang},\ and\ \citenamefont
  {Song}}]{Hu2024geometric}%
  \BibitemOpen
  \bibfield  {author} {\bibinfo {author} {\bibfnamefont {Y.~M.}\ \bibnamefont
  {Hu}}, \bibinfo {author} {\bibfnamefont {H.~Y.}\ \bibnamefont {Wang}},
  \bibinfo {author} {\bibfnamefont {Z.}~\bibnamefont {Wang}},\ and\ \bibinfo
  {author} {\bibfnamefont {F.}~\bibnamefont {Song}},\ }\href
  {https://doi.org/10.1103/PhysRevLett.132.050402} {\bibfield  {journal}
  {\bibinfo  {journal} {Phys. Rev. Lett.}\ }\textbf {\bibinfo {volume} {132}},\
  \bibinfo {pages} {050402} (\bibinfo {year} {2024})}\BibitemShut {NoStop}%
\bibitem [{\citenamefont {Xiao}\ \emph {et~al.}(2021)\citenamefont {Xiao},
  \citenamefont {Deng}, \citenamefont {Wang}, \citenamefont {Wang},
  \citenamefont {Yi},\ and\ \citenamefont {Xue}}]{Xiao2021observation}%
  \BibitemOpen
  \bibfield  {author} {\bibinfo {author} {\bibfnamefont {L.}~\bibnamefont
  {Xiao}}, \bibinfo {author} {\bibfnamefont {T.}~\bibnamefont {Deng}}, \bibinfo
  {author} {\bibfnamefont {K.}~\bibnamefont {Wang}}, \bibinfo {author}
  {\bibfnamefont {Z.}~\bibnamefont {Wang}}, \bibinfo {author} {\bibfnamefont
  {W.}~\bibnamefont {Yi}},\ and\ \bibinfo {author} {\bibfnamefont
  {P.}~\bibnamefont {Xue}},\ }\href
  {https://doi.org/10.1103/PhysRevLett.126.230402} {\bibfield  {journal}
  {\bibinfo  {journal} {Phys. Rev. Lett.}\ }\textbf {\bibinfo {volume} {126}},\
  \bibinfo {pages} {230402} (\bibinfo {year} {2021})}\BibitemShut {NoStop}%
\bibitem [{\citenamefont {Hu}\ \emph {et~al.}(2023)\citenamefont {Hu},
  \citenamefont {Zhang}, \citenamefont {Yue}, \citenamefont {Liao},
  \citenamefont {Liu}, \citenamefont {Zhang}, \citenamefont {Cheng},
  \citenamefont {Liu},\ and\ \citenamefont {Christensen}}]{Hu2023anti}%
  \BibitemOpen
  \bibfield  {author} {\bibinfo {author} {\bibfnamefont {B.}~\bibnamefont
  {Hu}}, \bibinfo {author} {\bibfnamefont {Z.}~\bibnamefont {Zhang}}, \bibinfo
  {author} {\bibfnamefont {Z.}~\bibnamefont {Yue}}, \bibinfo {author}
  {\bibfnamefont {D.}~\bibnamefont {Liao}}, \bibinfo {author} {\bibfnamefont
  {Y.}~\bibnamefont {Liu}}, \bibinfo {author} {\bibfnamefont {H.}~\bibnamefont
  {Zhang}}, \bibinfo {author} {\bibfnamefont {Y.}~\bibnamefont {Cheng}},
  \bibinfo {author} {\bibfnamefont {X.}~\bibnamefont {Liu}},\ and\ \bibinfo
  {author} {\bibfnamefont {J.}~\bibnamefont {Christensen}},\ }\href
  {https://doi.org/10.1103/PhysRevLett.131.066601} {\bibfield  {journal}
  {\bibinfo  {journal} {Phys. Rev. Lett.}\ }\textbf {\bibinfo {volume} {131}},\
  \bibinfo {pages} {066601} (\bibinfo {year} {2023})}\BibitemShut {NoStop}%
\bibitem [{\citenamefont {Hatano}\ and\ \citenamefont
  {Nelson}(1996)}]{Hatano1996localization}%
  \BibitemOpen
  \bibfield  {author} {\bibinfo {author} {\bibfnamefont {N.}~\bibnamefont
  {Hatano}}\ and\ \bibinfo {author} {\bibfnamefont {D.~R.}\ \bibnamefont
  {Nelson}},\ }\href {https://doi.org/10.1103/PhysRevLett.77.570} {\bibfield
  {journal} {\bibinfo  {journal} {Phys. Rev. Lett.}\ }\textbf {\bibinfo
  {volume} {77}},\ \bibinfo {pages} {570} (\bibinfo {year} {1996})}\BibitemShut
  {NoStop}%
\bibitem [{sm()}]{sm}%
  \BibitemOpen
  \href@noop {} {\bibinfo  {journal} {See the details of derivations all over
  the manuscript and the two-dimensional model in the supplementary materials}\
  }\BibitemShut {NoStop}%
\bibitem [{\citenamefont {Nakamura}\ \emph {et~al.}(2024)\citenamefont
  {Nakamura}, \citenamefont {Bessho},\ and\ \citenamefont
  {Sato}}]{Nakamura2024Bulk}%
  \BibitemOpen
\bibfield  {journal} {  }\bibfield  {author} {\bibinfo {author} {\bibfnamefont
  {D.}~\bibnamefont {Nakamura}}, \bibinfo {author} {\bibfnamefont
  {T.}~\bibnamefont {Bessho}},\ and\ \bibinfo {author} {\bibfnamefont
  {M.}~\bibnamefont {Sato}},\ }\href
  {https://doi.org/10.1103/PhysRevLett.128.157601} {\bibfield  {journal}
  {\bibinfo  {journal} {Phys. Rev. Lett.}\ }\textbf {\bibinfo {volume} {132}},\
  \bibinfo {pages} {136401} (\bibinfo {year} {2024})}\BibitemShut {NoStop}%
\bibitem [{\citenamefont {Yokomizo}\ and\ \citenamefont
  {Murakami}(2019)}]{Yokomizo2019non}%
  \BibitemOpen
  \bibfield  {author} {\bibinfo {author} {\bibfnamefont {K.}~\bibnamefont
  {Yokomizo}}\ and\ \bibinfo {author} {\bibfnamefont {S.}~\bibnamefont
  {Murakami}},\ }\href {https://doi.org/10.1103/PhysRevLett.123.066404}
  {\bibfield  {journal} {\bibinfo  {journal} {Phys. Rev. Lett.}\ }\textbf
  {\bibinfo {volume} {123}},\ \bibinfo {pages} {066404} (\bibinfo {year}
  {2019})}\BibitemShut {NoStop}%
\bibitem [{\citenamefont {Huang}\ \emph {et~al.}(2016)\citenamefont {Huang},
  \citenamefont {Meng}, \citenamefont {Wang}, \citenamefont {Peng},
  \citenamefont {Zhang}, \citenamefont {Chen}, \citenamefont {Li},
  \citenamefont {Zhou},\ and\ \citenamefont {Zhang}}]{huang2016experimental}%
  \BibitemOpen
  \bibfield  {author} {\bibinfo {author} {\bibfnamefont {L.}~\bibnamefont
  {Huang}}, \bibinfo {author} {\bibfnamefont {Z.}~\bibnamefont {Meng}},
  \bibinfo {author} {\bibfnamefont {P.}~\bibnamefont {Wang}}, \bibinfo {author}
  {\bibfnamefont {P.}~\bibnamefont {Peng}}, \bibinfo {author} {\bibfnamefont
  {S.~L.}\ \bibnamefont {Zhang}}, \bibinfo {author} {\bibfnamefont
  {L.}~\bibnamefont {Chen}}, \bibinfo {author} {\bibfnamefont {D.}~\bibnamefont
  {Li}}, \bibinfo {author} {\bibfnamefont {Q.}~\bibnamefont {Zhou}},\ and\
  \bibinfo {author} {\bibfnamefont {J.}~\bibnamefont {Zhang}},\ }\href
  {https://doi.org/10.1038/nphys3672} {\bibfield  {journal} {\bibinfo
  {journal} {Nat. Phys.}\ }\textbf {\bibinfo {volume} {12}},\ \bibinfo {pages}
  {540} (\bibinfo {year} {2016})}\BibitemShut {NoStop}%
\bibitem [{\citenamefont {Okuma}\ and\ \citenamefont
  {Sato}(2019)}]{okuma2019topological}%
  \BibitemOpen
  \bibfield  {author} {\bibinfo {author} {\bibfnamefont {N.}~\bibnamefont
  {Okuma}}\ and\ \bibinfo {author} {\bibfnamefont {M.}~\bibnamefont {Sato}},\
  }\href {https://doi.org/10.1103/PhysRevLett.123.097701} {\bibfield  {journal}
  {\bibinfo  {journal} {Phys. Rev. Lett.}\ }\textbf {\bibinfo {volume} {123}},\
  \bibinfo {pages} {097701} (\bibinfo {year} {2019})}\BibitemShut {NoStop}%
\bibitem [{\citenamefont {Guo}\ \emph {et~al.}(2021{\natexlab{a}})\citenamefont
  {Guo}, \citenamefont {Liu}, \citenamefont {Zhao}, \citenamefont {Liu},\ and\
  \citenamefont {Chen}}]{guo2021exact}%
  \BibitemOpen
  \bibfield  {author} {\bibinfo {author} {\bibfnamefont {C.~X.}\ \bibnamefont
  {Guo}}, \bibinfo {author} {\bibfnamefont {C.~H.}\ \bibnamefont {Liu}},
  \bibinfo {author} {\bibfnamefont {X.~M.}\ \bibnamefont {Zhao}}, \bibinfo
  {author} {\bibfnamefont {Y.}~\bibnamefont {Liu}},\ and\ \bibinfo {author}
  {\bibfnamefont {S.}~\bibnamefont {Chen}},\ }\href
  {https://doi.org/10.1103/PhysRevLett.127.116801} {\bibfield  {journal}
  {\bibinfo  {journal} {Phys. Rev. Lett.}\ }\textbf {\bibinfo {volume} {127}},\
  \bibinfo {pages} {116801} (\bibinfo {year} {2021}{\natexlab{a}})}\BibitemShut
  {NoStop}%
\bibitem [{\citenamefont {Wang}\ \emph {et~al.}(2012)\citenamefont {Wang},
  \citenamefont {Yu}, \citenamefont {Fu}, \citenamefont {Miao}, \citenamefont
  {Huang}, \citenamefont {Chai}, \citenamefont {Zhai},\ and\ \citenamefont
  {Zhang}}]{Wang2012spin}%
  \BibitemOpen
  \bibfield  {author} {\bibinfo {author} {\bibfnamefont {P.}~\bibnamefont
  {Wang}}, \bibinfo {author} {\bibfnamefont {Z.~Q.}\ \bibnamefont {Yu}},
  \bibinfo {author} {\bibfnamefont {Z.}~\bibnamefont {Fu}}, \bibinfo {author}
  {\bibfnamefont {J.}~\bibnamefont {Miao}}, \bibinfo {author} {\bibfnamefont
  {L.}~\bibnamefont {Huang}}, \bibinfo {author} {\bibfnamefont
  {S.}~\bibnamefont {Chai}}, \bibinfo {author} {\bibfnamefont {H.}~\bibnamefont
  {Zhai}},\ and\ \bibinfo {author} {\bibfnamefont {J.}~\bibnamefont {Zhang}},\
  }\href {https://doi.org/10.1103/PhysRevLett.109.095301} {\bibfield  {journal}
  {\bibinfo  {journal} {Phys. Rev. Lett.}\ }\textbf {\bibinfo {volume} {109}},\
  \bibinfo {pages} {095301} (\bibinfo {year} {2012})}\BibitemShut {NoStop}%
\bibitem [{\citenamefont {Yang}\ \emph {et~al.}(2019)\citenamefont {Yang},
  \citenamefont {Peng}, \citenamefont {Zhu}, \citenamefont {Buljan},
  \citenamefont {Joannopoulos}, \citenamefont {Zhen},\ and\ \citenamefont
  {Solja{\v{c}}i{\'c}}}]{yang2019synthesis}%
  \BibitemOpen
  \bibfield  {author} {\bibinfo {author} {\bibfnamefont {Y.}~\bibnamefont
  {Yang}}, \bibinfo {author} {\bibfnamefont {C.}~\bibnamefont {Peng}}, \bibinfo
  {author} {\bibfnamefont {D.}~\bibnamefont {Zhu}}, \bibinfo {author}
  {\bibfnamefont {H.}~\bibnamefont {Buljan}}, \bibinfo {author} {\bibfnamefont
  {J.~D.}\ \bibnamefont {Joannopoulos}}, \bibinfo {author} {\bibfnamefont
  {B.}~\bibnamefont {Zhen}},\ and\ \bibinfo {author} {\bibfnamefont
  {M.}~\bibnamefont {Solja{\v{c}}i{\'c}}},\ }\href
  {https://doi.org/10.1126/science.aay3183} {\bibfield  {journal} {\bibinfo
  {journal} {Science}\ }\textbf {\bibinfo {volume} {365}},\ \bibinfo {pages}
  {1021} (\bibinfo {year} {2019})}\BibitemShut {NoStop}%
\bibitem [{\citenamefont {Guo}\ \emph {et~al.}(2021{\natexlab{b}})\citenamefont
  {Guo}, \citenamefont {Jiang}, \citenamefont {Zhang}, \citenamefont {Zhang},
  \citenamefont {Zhang}, \citenamefont {Yang}, \citenamefont {Zhang},\ and\
  \citenamefont {Chan}}]{guo2021experimental}%
  \BibitemOpen
  \bibfield  {author} {\bibinfo {author} {\bibfnamefont {Q.}~\bibnamefont
  {Guo}}, \bibinfo {author} {\bibfnamefont {T.}~\bibnamefont {Jiang}}, \bibinfo
  {author} {\bibfnamefont {R.~Y.}\ \bibnamefont {Zhang}}, \bibinfo {author}
  {\bibfnamefont {L.}~\bibnamefont {Zhang}}, \bibinfo {author} {\bibfnamefont
  {Z.-Q.}\ \bibnamefont {Zhang}}, \bibinfo {author} {\bibfnamefont
  {B.}~\bibnamefont {Yang}}, \bibinfo {author} {\bibfnamefont {S.}~\bibnamefont
  {Zhang}},\ and\ \bibinfo {author} {\bibfnamefont {C.~T.}\ \bibnamefont
  {Chan}},\ }\href {https://doi.org/10.1038/s41586-021-03521-3} {\bibfield
  {journal} {\bibinfo  {journal} {Nature}\ }\textbf {\bibinfo {volume} {594}},\
  \bibinfo {pages} {195} (\bibinfo {year} {2021}{\natexlab{b}})}\BibitemShut
  {NoStop}%
\bibitem [{\citenamefont {Sun}\ \emph {et~al.}(2022)\citenamefont {Sun},
  \citenamefont {Zhang}, \citenamefont {Yu}, \citenamefont {Tian},
  \citenamefont {Chen},\ and\ \citenamefont {Sun}}]{sun2022non}%
  \BibitemOpen
  \bibfield  {author} {\bibinfo {author} {\bibfnamefont {Y.~K.}\ \bibnamefont
  {Sun}}, \bibinfo {author} {\bibfnamefont {X.~L.}\ \bibnamefont {Zhang}},
  \bibinfo {author} {\bibfnamefont {F.}~\bibnamefont {Yu}}, \bibinfo {author}
  {\bibfnamefont {Z.~N.}\ \bibnamefont {Tian}}, \bibinfo {author}
  {\bibfnamefont {Q.~D.}\ \bibnamefont {Chen}},\ and\ \bibinfo {author}
  {\bibfnamefont {H.~B.}\ \bibnamefont {Sun}},\ }\href
  {https://doi.org/10.1038/s41567-022-01669-x} {\bibfield  {journal} {\bibinfo
  {journal} {Nat. Phys.}\ }\textbf {\bibinfo {volume} {18}},\ \bibinfo {pages}
  {1080} (\bibinfo {year} {2022})}\BibitemShut {NoStop}%
\bibitem [{\citenamefont {Zhang}\ \emph
  {et~al.}(2022{\natexlab{b}})\citenamefont {Zhang}, \citenamefont {Yu},
  \citenamefont {Chen}, \citenamefont {Tian}, \citenamefont {Chen},
  \citenamefont {Sun},\ and\ \citenamefont {Ma}}]{zhang2022non}%
  \BibitemOpen
  \bibfield  {author} {\bibinfo {author} {\bibfnamefont {X.~L.}\ \bibnamefont
  {Zhang}}, \bibinfo {author} {\bibfnamefont {F.}~\bibnamefont {Yu}}, \bibinfo
  {author} {\bibfnamefont {Z.~G.}\ \bibnamefont {Chen}}, \bibinfo {author}
  {\bibfnamefont {Z.~N.}\ \bibnamefont {Tian}}, \bibinfo {author}
  {\bibfnamefont {Q.~D.}\ \bibnamefont {Chen}}, \bibinfo {author}
  {\bibfnamefont {H.~B.}\ \bibnamefont {Sun}},\ and\ \bibinfo {author}
  {\bibfnamefont {G.}~\bibnamefont {Ma}},\ }\href
  {https://doi.org/10.1038/s41566-022-00976-2} {\bibfield  {journal} {\bibinfo
  {journal} {Nat. Photonics}\ }\textbf {\bibinfo {volume} {16}},\ \bibinfo
  {pages} {390} (\bibinfo {year} {2022}{\natexlab{b}})}\BibitemShut {NoStop}%
\bibitem [{\citenamefont {Cheng}\ \emph {et~al.}(2023)\citenamefont {Cheng},
  \citenamefont {Wang},\ and\ \citenamefont {Fan}}]{cheng2023artificial}%
  \BibitemOpen
  \bibfield  {author} {\bibinfo {author} {\bibfnamefont {D.}~\bibnamefont
  {Cheng}}, \bibinfo {author} {\bibfnamefont {K.}~\bibnamefont {Wang}},\ and\
  \bibinfo {author} {\bibfnamefont {S.}~\bibnamefont {Fan}},\ }\href
  {https://doi.org/10.1103/PhysRevLett.130.083601} {\bibfield  {journal}
  {\bibinfo  {journal} {Phys. Rev. Lett.}\ }\textbf {\bibinfo {volume} {130}},\
  \bibinfo {pages} {083601} (\bibinfo {year} {2023})}\BibitemShut {NoStop}%
\bibitem [{\citenamefont {Pang}\ \emph {et~al.}(2024)\citenamefont {Pang},
  \citenamefont {Wong}, \citenamefont {Hu},\ and\ \citenamefont
  {Yang}}]{pang2024synthetic}%
  \BibitemOpen
  \bibfield  {author} {\bibinfo {author} {\bibfnamefont {Z.}~\bibnamefont
  {Pang}}, \bibinfo {author} {\bibfnamefont {B.~T.~T.}\ \bibnamefont {Wong}},
  \bibinfo {author} {\bibfnamefont {J.}~\bibnamefont {Hu}},\ and\ \bibinfo
  {author} {\bibfnamefont {Y.}~\bibnamefont {Yang}},\ }\href
  {https://doi.org/10.1103/PhysRevLett.132.043804} {\bibfield  {journal}
  {\bibinfo  {journal} {Phys. Rev. Lett.}\ }\textbf {\bibinfo {volume} {132}},\
  \bibinfo {pages} {043804} (\bibinfo {year} {2024})}\BibitemShut {NoStop}%
\bibitem [{\citenamefont {Xu}\ and\ \citenamefont {Zhang}(2016)}]{Xu2016dirac}%
  \BibitemOpen
  \bibfield  {author} {\bibinfo {author} {\bibfnamefont {Y.}~\bibnamefont
  {Xu}}\ and\ \bibinfo {author} {\bibfnamefont {C.}~\bibnamefont {Zhang}},\
  }\href {https://doi.org/10.1103/PhysRevA.93.063606} {\bibfield  {journal}
  {\bibinfo  {journal} {Phys. Rev. A}\ }\textbf {\bibinfo {volume} {93}},\
  \bibinfo {pages} {063606} (\bibinfo {year} {2016})}\BibitemShut {NoStop}%
\bibitem [{\citenamefont {Lin}\ \emph {et~al.}(2011)\citenamefont {Lin},
  \citenamefont {Jim{\'e}nez-Garc{\'\i}a},\ and\ \citenamefont
  {Spielman}}]{lin2011spin}%
  \BibitemOpen
  \bibfield  {author} {\bibinfo {author} {\bibfnamefont {Y.~J.}\ \bibnamefont
  {Lin}}, \bibinfo {author} {\bibfnamefont {K.}~\bibnamefont
  {Jim{\'e}nez-Garc{\'\i}a}},\ and\ \bibinfo {author} {\bibfnamefont {I.~B.}\
  \bibnamefont {Spielman}},\ }\href {https://doi.org/10.1038/nature09887}
  {\bibfield  {journal} {\bibinfo  {journal} {Nature}\ }\textbf {\bibinfo
  {volume} {471}},\ \bibinfo {pages} {83} (\bibinfo {year} {2011})}\BibitemShut
  {NoStop}%
\bibitem [{\citenamefont {Cheng}\ \emph {et~al.}(2024)\citenamefont {Cheng},
  \citenamefont {Wang}, \citenamefont {Roques-Carmes}, \citenamefont {Lustig},
  \citenamefont {Long}, \citenamefont {Wang},\ and\ \citenamefont
  {Fan}}]{Cheng2024Non}%
  \BibitemOpen
  \bibfield  {author} {\bibinfo {author} {\bibfnamefont {D.}~\bibnamefont
  {Cheng}}, \bibinfo {author} {\bibfnamefont {K.}~\bibnamefont {Wang}},
  \bibinfo {author} {\bibfnamefont {C.}~\bibnamefont {Roques-Carmes}}, \bibinfo
  {author} {\bibfnamefont {E.}~\bibnamefont {Lustig}}, \bibinfo {author}
  {\bibfnamefont {O.~Y.}\ \bibnamefont {Long}}, \bibinfo {author}
  {\bibfnamefont {H.}~\bibnamefont {Wang}},\ and\ \bibinfo {author}
  {\bibfnamefont {S.}~\bibnamefont {Fan}},\ }\href
  {https://doi.org/10.48550/arXiv.2406.00321} {\bibfield  {journal} {\bibinfo
  {journal} {arXiv preprint arXiv:2406.00321}\ } (\bibinfo {year}
  {2024})}\BibitemShut {NoStop}%
\bibitem [{\citenamefont {Okuma}\ \emph {et~al.}(2020)\citenamefont {Okuma},
  \citenamefont {Kawabata}, \citenamefont {Shiozaki},\ and\ \citenamefont
  {Sato}}]{Okuma2020}%
  \BibitemOpen
  \bibfield  {author} {\bibinfo {author} {\bibfnamefont {N.}~\bibnamefont
  {Okuma}}, \bibinfo {author} {\bibfnamefont {K.}~\bibnamefont {Kawabata}},
  \bibinfo {author} {\bibfnamefont {K.}~\bibnamefont {Shiozaki}},\ and\
  \bibinfo {author} {\bibfnamefont {M.}~\bibnamefont {Sato}},\ }\href
  {https://doi.org/10.1103/PhysRevLett.124.086801} {\bibfield  {journal}
  {\bibinfo  {journal} {Phys. Rev. Lett.}\ }\textbf {\bibinfo {volume} {124}},\
  \bibinfo {pages} {086801} (\bibinfo {year} {2020})}\BibitemShut {NoStop}%
\bibitem [{\citenamefont {Choi}\ \emph {et~al.}(2018)\citenamefont {Choi},
  \citenamefont {Hahn}, \citenamefont {Yoon},\ and\ \citenamefont
  {Song}}]{Choi2018}%
  \BibitemOpen
  \bibfield  {author} {\bibinfo {author} {\bibfnamefont {Y.}~\bibnamefont
  {Choi}}, \bibinfo {author} {\bibfnamefont {C.}~\bibnamefont {Hahn}}, \bibinfo
  {author} {\bibfnamefont {J.~W.}\ \bibnamefont {Yoon}},\ and\ \bibinfo
  {author} {\bibfnamefont {S.~H.}\ \bibnamefont {Song}},\ }\href
  {https://doi.org/10.1038/s41467-018-04690-y} {\bibfield  {journal} {\bibinfo
  {journal} {Nat. Commun.}\ }\textbf {\bibinfo {volume} {9}},\ \bibinfo {pages}
  {2182} (\bibinfo {year} {2018})}\BibitemShut {NoStop}%
\bibitem [{\citenamefont {Kato}\ \emph {et~al.}(2019)\citenamefont {Kato},
  \citenamefont {N{\'e}met}, \citenamefont {Senga}, \citenamefont {Mizukami},
  \citenamefont {Huang}, \citenamefont {Parkins},\ and\ \citenamefont
  {Aoki}}]{kato2019}%
  \BibitemOpen
  \bibfield  {author} {\bibinfo {author} {\bibfnamefont {S.}~\bibnamefont
  {Kato}}, \bibinfo {author} {\bibfnamefont {N.}~\bibnamefont {N{\'e}met}},
  \bibinfo {author} {\bibfnamefont {K.}~\bibnamefont {Senga}}, \bibinfo
  {author} {\bibfnamefont {S.}~\bibnamefont {Mizukami}}, \bibinfo {author}
  {\bibfnamefont {X.}~\bibnamefont {Huang}}, \bibinfo {author} {\bibfnamefont
  {S.}~\bibnamefont {Parkins}},\ and\ \bibinfo {author} {\bibfnamefont
  {T.}~\bibnamefont {Aoki}},\ }\href
  {https://doi.org/10.1038/s41467-019-08975-8} {\bibfield  {journal} {\bibinfo
  {journal} {Nat. Commun.}\ }\textbf {\bibinfo {volume} {10}},\ \bibinfo
  {pages} {1160} (\bibinfo {year} {2019})}\BibitemShut {NoStop}%
\bibitem [{\citenamefont {Fu}\ \emph {et~al.}(2021)\citenamefont {Fu},
  \citenamefont {Hu},\ and\ \citenamefont {Wan}}]{Fu2021}%
  \BibitemOpen
  \bibfield  {author} {\bibinfo {author} {\bibfnamefont {Y.}~\bibnamefont
  {Fu}}, \bibinfo {author} {\bibfnamefont {J.}~\bibnamefont {Hu}},\ and\
  \bibinfo {author} {\bibfnamefont {S.}~\bibnamefont {Wan}},\ }\href
  {https://doi.org/10.1103/PhysRevB.103.045420} {\bibfield  {journal} {\bibinfo
   {journal} {Phys. Rev. B}\ }\textbf {\bibinfo {volume} {103}},\ \bibinfo
  {pages} {045420} (\bibinfo {year} {2021})}\BibitemShut {NoStop}%
\end{thebibliography}%

\onecolumngrid
%\appendix

\renewcommand\thefigure{S\arabic{figure}}
\renewcommand\thetable{S\arabic{table}}
\renewcommand{\theequation}{S\arabic{equation}}

\section{The exact solution of the system without magnetic field}
\setcounter{figure}{0} 
\setcounter{equation}{0} 

We consider the Hamiltonian without magnetic field 
\begin{equation}
	H_z=\sum_{i}\left( 
	\begin{array}{cc}
		c_{i,\mathbf{+}}^{\dag } & c_{i,\mathbf{-}}^{\dag }%
	\end{array}%
	\right) t_{1}e^{i\theta \sigma _{z}}\left( 
	\begin{array}{c}
		c_{i+1,\mathbf{+}} \\ 
		c_{i+1,\mathbf{-}}%
	\end{array}%
	\right) +\left( 
	\begin{array}{cc}
		c_{i+1,\mathbf{+}}^{\dag } & c_{i+1,\mathbf{-}}^{\dag }%
	\end{array}%
	\right) t_{2}e^{i\phi \sigma _{z}}\left( 
	\begin{array}{c}
		c_{i,\mathbf{+}} \\ 
		c_{i,\mathbf{-}}%
	\end{array}%
	\right) .
\end{equation}%
The PBC eigenenergy can be expressed as 
\begin{equation}\label{eq2}
	E_{\pm }(q)=t_{1}e^{\pm i\theta }e^{iq}+t_{2}e^{\pm i\phi
	}e^{-iq},  
\end{equation}%
where $q$ varies from $0$ to $2\pi $. The lattice constant is assumed to be 1. 
For the system in OBC, the Hamiltonian can be diagonalized into two blocks
due to the fact that $\left[ \sigma _{z},H_{z}\right] =0$. The two blocks
are classified by the spin quantum number $\pm $ 
\begin{equation*}
	\left( 
	\begin{array}{cccccc}
		0 & e^{i\theta } & 0 & 0 &  &  \\ 
		e^{i\phi } & 0 & e^{i\theta } & 0 &  &  \\ 
		0 & e^{i\phi } & 0 & e^{i\theta } &  &  \\ 
		0 & 0 & e^{i\phi } & \ldots  &  &  \\ 
		&  &  &  & 0 & e^{i\theta } \\ 
		&  &  &  & e^{i\phi } & 0%
	\end{array}%
	\right) ,\left( 
	\begin{array}{cccccc}
		0 & e^{-i\theta } & 0 & 0 &  &  \\ 
		e^{-i\phi } & 0 & e^{-i\theta } & 0 &  &  \\ 
		0 & e^{-i\phi } & 0 & e^{-i\theta } &  &  \\ 
		0 & 0 & e^{-i\phi } & \ldots  &  &  \\ 
		&  &  &  & 0 & e^{-i\theta } \\ 
		&  &  &  & e^{-i\phi } & 0%
	\end{array}%
	\right) .
\end{equation*}%
The eigenvalues of the two matrices, $E_{\pm }\left( j=1,\ldots
N\right) $, are  
\begin{equation}
	E_{\pm }\left( j\right) =2\sqrt{e^{\pm i\theta }e^{\pm i\phi }}\cos  
	\frac{j\pi }{N+1} .
\end{equation}%
Consequently, the right eigenvectors are 
given by%
\begin{equation}
	\left\vert \psi _{\pm }\left( j\right) \right\rangle =\sum_{k=1}^{N}\left( 
	\sqrt{e^{\pm i\left( \theta -\phi \right) }}\right) ^{k}\sin \frac{ k j\pi }{%
		N+1}\left\vert k,\pm \right\rangle.
\end{equation}%
The eigenvectors are not normalized and $\left\vert k, \pm \right\rangle$ are the eigenstates of $\sigma _{z}$ at site $k$. 

\section{Minimum model}

\label{Appendix_b}

\subsection{One-dimensional model}

We consider the non-hermitian spin chain 
\begin{eqnarray}
	H &=&\sum_{i}\left( 
	\begin{array}{cc}
		c_{i,\mathbf{+}}^{\dag } & c_{i,\mathbf{-}}^{\dag }%
	\end{array}
	\right) t_{1}\left( 
	\begin{array}{cc}
		e^{i\theta } & 0 \\ 
		0 & e^{-i\theta }%
	\end{array}
	\right) \left( 
	\begin{array}{c}
		c_{i+1,\mathbf{+}} \\ 
		c_{i+1,\mathbf{-}}%
	\end{array}
	\right) +\left( 
	\begin{array}{cc}
		c_{i+1,\mathbf{+}}^{\dag } & c_{i+1,\mathbf{-}}^{\dag }%
	\end{array}
	\right) t_{2}\left( 
	\begin{array}{cc}
		e^{i\phi } & 0 \\

		0 & -e^{-i\phi }%
	\end{array}
	\right) \left( 
	\begin{array}{c}
		c_{i,\mathbf{+}} \\ 
		c_{i,\mathbf{-}}%
	\end{array}
	\right) .
\end{eqnarray}
Using the Fourier transform $c_{i}=\sum_{q}e^{iqR_i}c_{q}$, we rewrite the
Hamiltonian in momentum space 
\begin{eqnarray*}
	H &=&\sum_{{q}}\left( 
	\begin{array}{cc}
		c_{{q,+}}^{\dag } & c_{{q,-}}^{\dag }%
	\end{array}
	\right) \Lambda _{q}\left( 
	\begin{array}{c}
		c_{{q,+}} \\ 
		c_{{q,-}}%
	\end{array}
	\right) ,
\end{eqnarray*}
where 
\begin{equation*}
	\Lambda _{q}=\left( 
	\begin{array}{cc}
		t_{1}e^{i\theta }e^{iq     }+t_{2}e^{i\phi }e^{-i{q     }} & 0 \\ 
		0 & t_{1}e^{-i\theta }e^{i{q     }}+t_{2}e^{-i\phi }e^{-i{q     }}%
	\end{array}
	\right) .
\end{equation*}

When we diagonalize the matrix $\Lambda _{q}$, which is non-Hermitian, we
can find the two bands of the system.
If we write the Hamiltonian as 
\begin{equation*}
	H=\sum_{q}e^{iq     }\left( 
	\begin{array}{cc}
		c_{{q,+}}^{\dag } & c_{{q,-}}^{\dag }%
	\end{array}
	\right) \left( 
	\begin{array}{cc}
		a & 0 \\ 
		0 & \frac{1}{2}%
	\end{array}
	\right) \left( 
	\begin{array}{c}
		c_{{q,+}} \\ 
		c_{{q,-}}%
	\end{array}
	\right) +e^{-i{q    }}\left( 
	\begin{array}{cc}
		c_{{q,+}}^{\dag } & c_{{q,-}}^{\dag }%
	\end{array}
	\right) \left( 
	\begin{array}{cc}
		1 & 0 \\ 
		0 & \frac{b}{2}%
	\end{array}
	\right) \left( 
	\begin{array}{c}
		c_{{q,+}} \\ 
		c_{{q,-}}%
	\end{array}
	\right),
\end{equation*}
we obtain the relations 
\begin{eqnarray*}
	t_{1}^{2} &=&\frac{a}{2}, \\
	e^{2i\theta } &=&2a, \\
	t_{2}^{2} &=&\frac{b}{2}, \\
	e^{2i\phi } &=&\frac{2}{b}.
\end{eqnarray*}
Note that $\theta $ and $\phi $ must be imaginary if $a$ and $b$ are real. In real
space, the toy model is in fact two copies of the HN model 
\begin{equation}
	H=\sum_{i}ac_{i,+}^{\dag }c_{i+1,+}+c_{i+1,+}^{\dag }c_{i,+}+\frac{1}{2}
	c_{i,-}^{\dag }c_{i+1,-}+\frac{b}{2}c_{i+1,-}^{\dag }c_{i,-}.
\end{equation}
$\left|a\right|>1$ leads the spin-up states skinning to the left, and  $\left|b\right|>1$ makes the spin-down states localized to the right. Conversely, when $\left|a\right|<1$ and $\left|b\right|<1$, the direction of the NHSE for both spins is reversed. When $\left|a\right|=\left|b\right|\neq 1$, 
a symmetric bidirectional skin effect with spin separation appears.

Although it was recently demonstrated that simultaneous presence of non-Hermitian skin modes at both boundaries can emerge due to the effective next nearest neighbor (NNN) hopping induced by the non-Abelian SU(2) gauge fields \cite{pang2024synthetic}, its underlying physical mechanism is fundamentally different from our case, and the spin-up and spin-down states in these studies were not separated. It was also shown that $Z_2$ NHSE with spin separation to opposite boundaries can be realized in a stacked system with time-reversal symmetry inspired by the quantum spin Hall insulator \cite{Okuma2020}. In comparison, the NHSE with spin separation in our system does not require such symmetry protection.
\subsection{Two-dimensional model}

Similar to the 1D case, we consider the non-Hermitian 2D model
\begin{eqnarray}
	H &=&\sum_{i,j}\left( 
	\begin{array}{cc}
		c_{i,j\mathbf{,+}}^{\dag } & c_{i.j\mathbf{,-}}^{\dag }%
	\end{array}
	\right) t_{1}e^{i\theta \sigma _{z}}\left( 
	\begin{array}{c}
		c_{i+1,j,+} \\ 
		c_{i+1,j,-}%
	\end{array}
	\right) +\left( 
	\begin{array}{cc}
		c_{i+1,j,+}^{\dag } & c_{i+1,j,-}^{\dag }%
	\end{array}
	\right) t_{2}e^{i\phi \sigma _{z}}\left( 
	\begin{array}{c}
		c_{i,j,+} \\ 
		c_{i,j,+}%
	\end{array}
	\right)  \notag \\
	&&+\left( 
	\begin{array}{cc}
		c_{i,j\mathbf{,+}}^{\dag } & c_{i.j\mathbf{,-}}^{\dag }%
	\end{array}
	\right) t_{3}e^{i\theta' \sigma _{z}}\left( 
	\begin{array}{c}
		c_{i,j+1,+} \\ 
		c_{i,j+1,-}%
	\end{array}
	\right) +\left( 
	\begin{array}{cc}
		c_{i,j+1,+}^{\dag } & c_{i,j+1,-}^{\dag }%
	\end{array}
	\right) t_{4}e^{i\phi' \sigma _{z}}\left( 
	\begin{array}{c}
		c_{i,j,+} \\ 
		c_{i,j,+}%
	\end{array}
	\right) .
\end{eqnarray}
We also use the Fourier transform 
$c_{i,j}=\sum_{\mathbf{q}} e^{i({q}_{x}{R}_{i}+{q}_{y}{R}_{j})}c_{\mathbf{q}}$ to rewrite the Hamiltonian in momentum space: 
\begin{eqnarray}
	H &=&\sum_{\mathbf{q}}\left( 
	\begin{array}{cc}
		c_{\mathbf{q,+}}^{\dag } & c_{\mathbf{q,-}}^{\dag }%
	\end{array}
	\right) \Lambda _{\mathbf{q}}\left( 
	\begin{array}{c}
		c_{\mathbf{q,+}} \\ 
		c_{\mathbf{q,-}}%
	\end{array}
	\right) ,
\end{eqnarray}
where 
\begin{equation}
	\Lambda _{\mathbf{q}}=t_{1}e^{iq_{x}   }\left( 
	\begin{array}{cc}
		e^{i\theta } & 0 \\ 
		0 & e^{-i\theta }%
	\end{array}
	\right) +t_{2}e^{-iq_{x}   }\left( 
	\begin{array}{cc}
		e^{i\phi } & 0 \\ 
		0 & e^{-i\phi }%
	\end{array}
	\right) +t_{3}e^{iq_{y}   }\left( 
	\begin{array}{cc}
		e^{i\alpha } & 0 \\ 
		0 & e^{-i\alpha }%
	\end{array}
	\right) +t_{4}e^{-iq_{y}   }\left( 
	\begin{array}{cc}
		e^{i\beta } & 0 \\ 
		0 & e^{-i\beta }%
	\end{array}
	\right) ,
\end{equation}
with the lattice vectors equal to $(1,0)$ and $(0, 1)$ for simplicity.

If we write the Hamiltonian in the following form:
\begin{equation}
	H=\sum_{\mathbf{q}}\left( 
	\begin{array}{cc}
		c_{\mathbf{q,+}}^{\dag } & c_{\mathbf{q,-}}^{\dag }%
	\end{array}
	\right) \left[ e^{iq_{x}   }\left( 
	\begin{array}{cc}
		a & 0 \\ 
		0 & \frac{1}{2}%
	\end{array}
	\right) +e^{-iq_{x}  }\left( 
	\begin{array}{cc}
		1 & 0 \\ 
		0 & \frac{b}{2}%
	\end{array}
	\right) +e^{iq_{y}   }\left( 
	\begin{array}{cc}
		\tilde{a} & 0 \\ 
		0 & \frac{1}{2}%
	\end{array}
	\right) +e^{-iq_{y}   }\left( 
	\begin{array}{cc}
		1 & 0 \\ 
		0 & \frac{\tilde{b}}{2}%
	\end{array}
	\right) \right] \left( 
	\begin{array}{c}
		c_{\mathbf{q,+}} \\ 
		c_{\mathbf{q,-}}%
	\end{array}
	\right),
\end{equation}
we then obtain the following relations: 
\begin{eqnarray*}
	t_{1}e^{i\theta } &=&a, \\
	t_{1}e^{-i\theta } &=&\frac{1}{2}, \\
	t_{2}e^{i\phi } &=&1, \\
	t_{2}e^{-i\phi } &=&\frac{b}{2}, \\
	t_{3}e^{i\theta' } &=&\tilde{a}, \\
	t_{3}e^{-i\theta' } &=&\frac{1}{2}, \\
	t_{4}e^{i\phi' } &=&1, \\
	t_{4}e^{i\phi' } &=&\frac{\tilde{b}}{2}.
\end{eqnarray*}
In real space, we find similar results:
\begin{eqnarray}
	H &=&\sum_{i,j}ac_{i,j,+}^{\dag }c_{i+1,j,+}+c_{i+1,j,+}^{\dag }c_{i,j,+}+ 
	\frac{1}{2}c_{i,j,-}^{\dag }c_{i+1,j,-}+\frac{b}{2}c_{i+1,j,-}^{\dag
	}c_{i,j-}  \notag  \\
	&&+\sum_{i,j} \tilde{a} c_{i,j,+}^{\dag }c_{i,j+1,+}+c_{i,j+1,+}^{\dag }c_{i,j,+}+ 
	\frac{1}{2}c_{i,j,-}^{\dag }c_{i,j+1,-}+\frac{\tilde{b}}{2}c_{i,j+1,-}^{\dag
	}c_{i,j,-}.
\end{eqnarray}

\section{skin direction}
\subsection{Transition from bidirectional to unidirectional skin effect}

When the magnetic field is turned on, the PBC spectra are
\begin{eqnarray}
	E_{1,2} &=&\frac{1}{2} \left[\left( e^{iq}\left( e^{i\theta }+e^{-i\theta }\right)
	+e^{-iq}\left( e^{i\phi }+e^{-i\phi }\right) \right) \pm \sqrt{4\delta
		^{2}+\left( e^{iq}\left( e^{i\theta }-e^{-i\theta }\right) +e^{-iq}\left(
		e^{i\phi }-e^{-i\phi }\right) \right) ^{2}} \right], 
\end{eqnarray}
where $q$ is a wave vector. The lattice constant is set equal to 1. When $\delta $ is relatively large, $E_{1}$ and $E_{2}$ can be simplified using Taylor series:
\begin{eqnarray}
	E_{1,2} &=&\frac{1}{2} \left[ e^{iq}\left( e^{i\theta }+e^{-i\theta }\right)
	+e^{-iq}(e^{i\phi }+e^{-i\phi }) \right]\pm \frac{1}{8\delta}\left[ e^{iq}\left( e^{i\theta
	}-e^{-i\theta }\right) +e^{-iq}\left( e^{i\phi }-e^{-i\phi }\right) \right]
	^{2} \pm \delta. 
\end{eqnarray}
Thus, the winding direction depends on the
first term of $E_{1}$ and $E_{2}$, so that the sign of the winding number depends on the coefficients of $e^{iq}$ and $e^{-iq}$. By comparing $\left\vert \cosh(i\theta) \right\vert $ and $\left\vert
\cosh(i\phi) \right\vert $, we can obtain the winding number.

\subsection{Magnetically induced FOSE}
In the presence of the magnetic field, the PBC spectra with both $\theta$ and $\phi$ being real are
\begin{eqnarray}
	E_{1,2} &=& e^{iq} \cos\theta+e^{-iq}\cos\phi\pm\sqrt{\delta
		^{2}+\left(ie^{iq}\sin\theta+ie^{-iq}\sin\phi \right) ^{2}}   \notag \\
	&=& e^{iq}\cos\theta+e^{-iq}\cos\phi\pm\left(ie^{iq}\sin\theta+ie^{-iq}\sin\phi \right)\sqrt{1+\left(\frac{\delta
		}{ie^{iq}\sin\theta+ie^{-iq}\sin\phi}\right)^2 } . 
\end{eqnarray} 
When the magnetic field $\delta$ is weak, we have 
\begin{eqnarray}
	\lim_{\delta \to 0}{E_{1,2}} &=& e^{iq}\cos\theta+e^{-iq}\cos\phi\pm\left(ie^{iq}\sin\theta+ie^{-iq}\sin\phi \right)\pm\frac{1}{2}\frac{\delta^2}{ie^{iq}\sin\theta+ie^{-iq}\sin\phi}  \notag \\
	&=& e^{iq}e^{\pm i\theta}+e^{-iq}e^{\pm i\phi} \mp \frac{i}{2}\frac{\delta^2}{e^{iq}\sin\theta+e^{-iq}\sin\phi} .
\end{eqnarray} 
Note that $e^{iq}e^{\pm i\theta}+e^{-iq}e^{\pm i\phi}$ map to two straight lines in the complex plane. When $\delta$=0, the system is explicitly topologically trivial. In contrast, when $\delta \neq 0$, $H'=e^{iq}\sin\theta+e^{-iq}\sin\phi$ represents the Hamiltonian of the standard HN model with nonreciprocal hopping strengths for $|\sin\theta|\neq|\sin\phi|$. When $|\sin\theta|>|\sin\phi|$, an extensive number of OBC states are localized at one boundary. When $|\sin\theta|<|\sin\phi|$, many OBC states are localized at the other boundary. In other words, the magnetically induced term $\delta^2/(e^{iq}\sin\theta+e^{-iq}\sin\phi)$ forms a closed loop in the complex plane and leads to nonreciprocal effective hopping strengths which depend on phases $\theta$ and $\phi$. The system under an external magnetic field is therefore nontrivial with point-gap topology.

\subsection{Directional reversal of magnetically induced FOSE}

As the magnetic field amplitude increases, the winding number changes sign across specific lines. This indicates that our system is nontrivial. In other words, the topological property of our system is different before and after the point-gap reopens. In the presence of the magnetic field, the PBC spectra are
\begin{eqnarray}
	E_{1,2} &=& 2e^{iq}\cos \theta+2e^{-iq}\cos \phi\pm\sqrt{(2\delta)
		^{2}+\left(2ie^{iq}\sin \theta+2ie^{-iq}\sin \phi \right) ^{2}} .
\end{eqnarray} 
When $\phi=\theta$, $E_{1,2} = 4\cos \theta \cos q \pm4\sqrt{(\delta)^{2}-\left(\sin \theta \cos q \right) ^{2}}$.
Since $\cos q$ is an arc on the real axis, the image of $E_{1,2}(\cos q)$ will consist of arcs on the complex plane. In other words, both spectral areas are zero and the corresponding point-gaps disappear. For $\phi=\theta-\pi$, $\phi=-\theta$, and $\phi=-\theta+\pi$, the corresponding physical scenarios are the same. The underlying physical mechanism of the NHSE is associated with the asymmetric hopping strengths in the forward and backward directions. Thus, the point-gap topology collapses when the forward and backward hopping strengths are the same. We can consequently obtain some additional transition lines when 
\begin{eqnarray}
	\sqrt{(2\delta)^{2}+\left(2ie^{iq}\sin \theta+2ie^{-iq}\sin \phi \right) ^{2}}=\pm2ie^{iq}\sin \theta\pm2ie^{-iq}\sin \phi,  \\
	\sqrt{(2\delta)^{2}+\left(2ie^{iq}\sin \theta+2ie^{-iq}\sin \phi \right) ^{2}}=\pm2ie^{iq}\sin \theta\mp2ie^{-iq}\sin \phi.  
\end{eqnarray} 
The first equation requires zero magnetic field amplitude ($\delta$=0), which is inconsistent with our assumption that the system is under an external magnetic field. The second equation leads to $\delta=\pm 2\sqrt{\sin \theta \sin \phi}$, and the PBC spectra therefore collapse into arcs:
\begin{eqnarray}
	E_{1} &=& 2e^{iq}e^{\pm i\theta}+2e^{-iq}e^{\mp i\phi}, \\
	E_{2} &=& 2e^{iq}e^{\mp i\theta}+2e^{-iq}e^{\pm i\phi}.
\end{eqnarray} 

\section{Symmetry of eigenenergies and $\widetilde{\sigma }_{x,y}$ in the presence of a magnetic field}

The Hamiltonian in the presence of a magnetic field $\delta =\left(
\delta_x,0,0\right) $ is 
\begin{equation}
	H=\left( 
	\begin{array}{cccccccc}
		& \delta_x & e^{i\theta } &  &  &  &  &  \\ 
		\delta_x &  &  & e^{-i\theta } &  &  &  &  \\ 
		e^{i\phi } &  &  & \delta_x & e^{i\theta } &  &  &  \\ 
		& e^{-i\phi } & \delta_x &  &  & \ldots  &  &  \\ 
		&  & e^{i\phi } &  &  &  & e^{i\theta } &  \\ 
		&  &  & \ldots  &  &  &  & e^{-i\theta } \\ 
		&  &  &  & e^{i\phi } &  &  & \delta_x \\ 
		&  &  &  &  & e^{-i\phi } & \delta_x & 
	\end{array}%
	\right) .
\end{equation}%
Its $i$-th eigenvalue equation is $H\psi _{i}=E_{i}\psi _{i}$, i.e. 
\begin{equation}
	\left( 
	\begin{array}{cccccccc}
		& \delta_x & e^{i\theta } &  &  &  &  &  \\ 
		\delta_x &  &  & e^{-i\theta } &  &  &  &  \\ 
		e^{i\phi } &  &  & \delta_x & e^{i\theta } &  &  &  \\ 
		& e^{-i\phi } & \delta_x &  &  & \ldots  &  &  \\ 
		&  & e^{i\phi } &  &  &  & e^{i\theta } &  \\ 
		&  &  & \ldots  &  &  &  & e^{-i\theta } \\ 
		&  &  &  & e^{i\phi } &  &  & \delta_x \\ 
		&  &  &  &  & e^{-i\phi } & \delta_x & 
	\end{array}%
	\right) \left( 
	\begin{array}{c}
		x_{1,+}^{i} \\ 
		x_{1,-}^{i} \\ 
		x_{2,+}^{i} \\ 
		x_{2,-}^{i} \\ 
		\ldots  \\ 
		x_{N,+}^{i} \\ 
		x_{N,-}^{i}%
	\end{array}%
	\right) =E_{i}\left( 
	\begin{array}{c}
		x_{1,+}^{i} \\ 
		x_{1,-}^{i} \\ 
		x_{2,+}^{i} \\ 
		x_{2,-}^{i} \\ 
		\ldots  \\ 
		x_{N,+}^{i} \\ 
		x_{N,-}^{i}%
	\end{array}%
	\right) .
\end{equation}
We assume that states $n,l$ have opposite eigenenergies:
\begin{eqnarray}
	H\psi _{n} &=&E_{n}\psi _{n}, \\
	H\psi _{l} &=&E_{l}\psi _{n}=-E_{n}\psi _{l}.
\end{eqnarray}%
The following equations are then satisfied:%
\begin{eqnarray}
	\delta_x x_{1,-s}^{n}+e^{is\theta }x_{2,s}^{n} &=&E_{n}x_{1,s}^{n}, \label{en1}\\
	e^{is\phi }x_{j-1,s}^{n}+\delta_x x_{j,-s}^{n}+e^{is\theta }x_{j+1,s}^{n}
	&=&E_{n}x_{j,s}^{n}, \label{en2}\\
	e^{is\phi }x_{N-1,s}^{n}+\delta_x x_{N,-s}^{n} &=&E_{n}x_{N,s}^{n}, \label{en3}
\end{eqnarray}
and%
\begin{eqnarray*}
	\delta_x x_{1,-s}^{l}+e^{is\theta }x_{2,s}^{l} &=&-E_{n}x_{1,s}^{l}, \\
	e^{is\phi }x_{j-1,s}^{l}+\delta_x x_{j,-s}^{l}+e^{is\theta }x_{j+1,s}^{l}
	&=&-E_{n}x_{j,s}^{l}, \\
	e^{is\phi }x_{N-1,s}^{l}+\delta_x x_{N,-s}^{l} &=&-E_{n}x_{N,s}^{l},
\end{eqnarray*}%
with $j\in \left[ 2,N-1\right]$ and $s=\pm 1$. By summing or subtracting these, we obtain 
\begin{eqnarray*}
	\delta_x\left( x_{1,-}^{n}+x_{1,-}^{l}\right) +e^{i\theta }\left(
	x_{2,+}^{n}+x_{2,+}^{l}\right)  &=&E_{n}\left(
	x_{1,+}^{n}-x_{1,+}^{l}\right),  \\
	\delta_x\left( x_{1,+}^{n}-x_{1,+}^{l}\right) +e^{-i\theta }\left(
	x_{2,-}^{n}-x_{2,-}^{l}\right)  &=&E_{n}\left(
	x_{1,-}^{n}+x_{1,-}^{l}\right),  \\
	e^{is\phi }\left( x_{j-1,s}^{n}-s\left( -1\right) ^{j-2}x_{j-1,s}^{l}\right)
	+\delta_x\left( x_{j,-s}^{n}+s\left( -1\right) ^{j-1}x_{j,-s}^{l}\right) &&\\
	+e^{is\theta }\left( x_{j+1,s}^{n}-s\left( -1\right)
	^{j}x_{j+1,s}^{l}\right)  &=& E_{n}\left( x_{j,s}^{n}-s\left( -1\right)
	^{j-1}x_{j,s}^{l}\right),  \\
	e^{is\phi }\left( x_{N-1,s}^{n}-s\left( -1\right) ^{N-2}x_{N-1,s}^{l}\right)
	+\delta_x\left( x_{N,-s}^{n}+s\left( -1\right) ^{N-1}x_{N,-s}^{l}\right) 
	&=&E_{n}\left( x_{N,s}^{n}-s\left( -1\right) ^{N-1}x_{N,s}^{l}\right) .
\end{eqnarray*}%
Once the coefficients in the two eigenvectors satisfy%
\begin{equation}
	x_{j,s}^{n}=s\left( -1\right) ^{j-1}x_{j,s}^{l},
\end{equation}%
all eigenvalue equations are satisfied. In fact, in the energy spectra any eigenenergy is in pair with
its opposite eigenenergy, as also verified by numerical results. This also implies that the system has the symmetry $\sigma_z H(\beta) \sigma_z=-H(\beta e^{i\pi})$ in the GBZ. The
spin polarizations $\widetilde{\sigma }_{x,y}$ of the pair states at any site $i$ are opposite, 
\begin{equation} \label{sigmaxy}
	\widetilde{\sigma }_{n,x}^{(j)}=-\widetilde{\sigma }_{l,x}^{(j)}, \quad \widetilde{\sigma }_{n,y}^{(j)}=-\widetilde{\sigma }_{l,y}^{(j)},
\end{equation}
since 
\begin{equation}
	\left( x_{i,s}^{n}\right) ^{\ast }x_{i,-s}^{n}=-\left( x_{i,s}^{l}\right)
	^{\ast }x_{i,-s}^{l}.
\end{equation}

If $\theta ,\phi \in \mathbf{R},$ we may consider another symmetry:%
\begin{eqnarray}
	H\psi _{n} &=&E_{n}\psi _{n}, \\
	H\psi _{l} &=&E_{l} \psi _{l}=E_{n}^{\ast }\psi _{l}.
\end{eqnarray}%
The eigenvalue equations can then be written as 
\begin{eqnarray*}
	\delta_x x_{1,-s}^{n}+e^{is\theta }x_{2,s}^{n} &=&E_{n}x_{1,s}^{n}, \\
	e^{is\phi }x_{j-1,s}^{n}+\delta_x x_{j,-s}^{n}+e^{is\theta }x_{j+1,s}^{n}
	&=&E_{n}x_{j,s}^{n}, \\
	e^{is\phi }x_{N-1,s}^{n}+\delta_x x_{N,-s}^{n} &=&E_{n}x_{N,s}^{n}, \\
	\delta_x x_{1,-s}^{l}+e^{is\theta }x_{2,s}^{l} &=&E_{n}^{\ast }x_{1,s}^{l}, \\
	e^{is\phi }x_{j-1,s}^{l}+\delta_x x_{j,-s}^{l}+e^{is\theta }x_{j+1,s}^{l}
	&=&E_{n}^{\ast }x_{j,s}^{l}, \\
	e^{is\phi }x_{N-1,s}^{l}+\delta_x x_{N,-s}^{l} &=&E_{n}^{\ast }x_{N,s}^{l}.
\end{eqnarray*}%
The complex conjugate of the equations corresponding to the $l$-th state results in the condition 
\begin{equation}
	\left( x_{j,s}^{l}\right) ^{\ast }=x_{j,-s}^{n}
\end{equation}%
to satisfy 
$
E_{n}^{\ast }=E_{l}.
$
This implies that the system respects pseudo-Hermiticity in the GBZ
\begin{equation}
	\sigma_{x,y} H(\beta) \sigma_{x,y}=[H(\beta^* )]^\dag.
	\label{PH}
\end{equation}%
We thus infer that the eigenenergies have four-fold symmetry on the complex plane, and that, if the lattice number of the chain is odd, then there are at least two real eigenenergies with opposite sign. We also verified this numerically. For the spin polarization, besides the symmetries shown in Eq. (\ref{sigmaxy}), we have an extra symmetry 
\begin{equation}\label{sigma_z}
	\widetilde{\sigma }_{n,z}^{(j)}=-\widetilde{\sigma }_{l,z}^{(j)}.
\end{equation}

\section{Size-dependent fragility to magnetic fields}

\begin{figure}[H]
	\centering
	\includegraphics[width=1.1\linewidth, trim=100 0 0 0, clip]{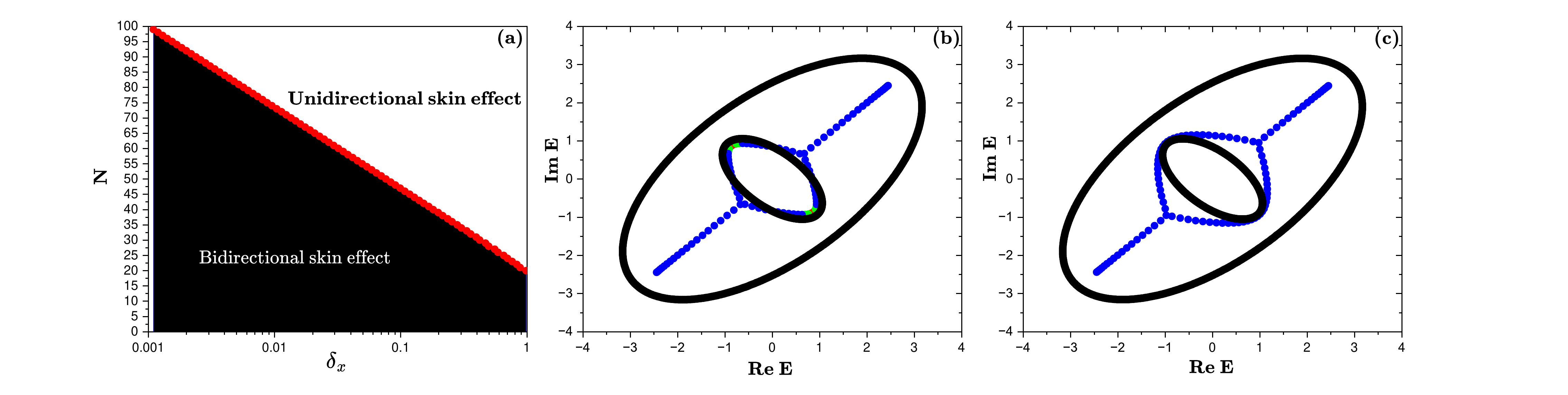}
	\caption{The system parameters are $\theta=\pi, e^{i \phi}=3 i$. (a) The relation between the required magnetic field $\Delta_x$ to polarize the skin effect and the lattice number. (b) The energy spectra in OBC and PBC for $N=50$ under a weak magnetic field $\delta_x=10^{-4}$. Some OBC states still remain in the smaller circle of the PBC energy spectra. (c) Under a magnetic field $\delta_x=0.1$ which is larger than the critical value $\Delta_x=0.0767$, all OBC states are in between the two PBC curves.} 
	\label{figs1}
\end{figure}

We consider the case that the PBC spectra form nested rings. Under a very weak magnetic field, the original bidirectional skin effect collapses to
one end of the chain, i.e. to a unidirectional skin effect. All OBC states are in between the two PBC energy curves. The critical magnetic field $\Delta_x$ to polarize the skin effect decreases exponentially with increasing lattice number, as shown in Fig. \ref{figs1}. As an example, for $\theta=\pi, e^{i \phi}=3 i$, we obtain 
$ \Delta_x \approx 0.983 \exp\left( -6.89 \frac{N-20}{80}\right).$

\section{$\mathcal{PT}$ symmetry and anti-$\mathcal{PT}$ symmetry}

\subsection{Energy Spectra}
When the magnetic field is large, the Hermitian Zeeman coupling dominates the system and can be treated as the unperturbed part of the Hamiltonian 
\begin{equation}
	H_{0}=\sum_{i} \delta_x c_{i}^{\dag } \sigma_x c_{i}.
\end{equation}
The other part of the Hamiltonian
\begin{equation}
	H_{1}=\sum_{i}c_{i}^{\dag }t_{1}e^{i\theta \mathbf{\sigma
			_z}}c_{i+1}+c_{i+1}^{\dag }t_{2}e^{i\phi \mathbf{\sigma_z}}c_{i}
\end{equation}
can then be regarded as the perturbation. We now assume that $\theta ,\phi \in \mathbf{R}$, and the unperturbed states
are represented by the eigenstates of $\sigma _{x},\left\vert \pm
\right\rangle =\frac{1}{\sqrt{2}}\left( 
\begin{array}{c}
	1 \\ 
	\pm 1%
\end{array}%
\right) ,$ so that
\begin{equation*}
	\left\vert i,\pm \right\rangle =\frac{1}{\sqrt{2}}\left( 
	\begin{array}{cccccc}
		\ldots & 0  & 1 & \pm 1 & 0  & \ldots
	\end{array}%
	\right) ^\intercal,
\end{equation*}
and the eigenenergies are degenerate: 
\begin{equation*}
	E_{i,\pm }=\pm \delta_x.
\end{equation*}%
In this basis, the matrix form of the perturbation is 
\begin{equation}
	H_{1}=\left( 
	\begin{array}{ccccccc}
		&  & e^{i\theta } &  &  &  &  \\ 
		&  &  & e^{-i\theta } &  &  &  \\ 
		e^{i\phi } &  &  &  & \ldots  &  &  \\ 
		& e^{-i\phi } &  &  &  & e^{i\theta } &  \\ 
		&  & \ldots  &  &  &  & e^{-i\theta } \\ 
		&  &  & e^{i\phi } &  &  &  \\ 
		&  &  &  & e^{-i\phi } &  & 
	\end{array}%
	\right)_{2N \times 2N} .
\end{equation}%
We use the degenerate perturbation theory. The two groups of eigenstates $\left\vert i,+\right\rangle $ and $\left\vert i,-\right\rangle $
form two subspaces of the Hilbert space. The first-order corrections of the degenerate energies are given by the eigenvalues of two matrices, which are in fact identical: 
\begin{equation}
	W_{\pm }=\left( 
	\begin{array}{cccccc}
		& \cos \theta  &  &  &  &  \\ 
		\cos \phi  &  & \cos \theta  &  &  &  \\ 
		& \cos \phi  &  & \ldots  &  &  \\ 
		&  & \ldots  &  &  &  \\ 
		&  &  &  &  & \cos \theta  \\ 
		&  &  &  & \cos \phi  & 
	\end{array}%
	\right) _{N\times N}.
\end{equation}%
The first-order corrections of the eigenenergies are 
\begin{equation}
	E_{j,\pm }^{\left( 1\right) }=2\sqrt{\cos \theta \cos \phi }\cos \left( 
	\frac{j\pi }{N+1}\right) ,
\end{equation}%
with corresponding eigenvectors%
\begin{equation*}
	\left( 
	\begin{array}{c}
		\sqrt{\frac{\cos \theta }{\cos \phi }}\sin \frac{j\pi }{N+1} \\ 
		\ldots  \\ 
		\left( \sqrt{\frac{\cos \theta }{\cos \phi }}\right)^{k}\sin \frac{kj\pi }{N+1} \\ 
		\ldots  \\ 
		\left( \sqrt{\frac{\cos \theta }{\cos \phi }}\right)^{N}\sin \frac{Nj\pi }{N+1}%
	\end{array}%
	\right) .
\end{equation*}%
Given that the magnetic field is strong, the two branches of eigenenergies
with corrections are not close to each other. In each subspace, the reconstructed
eigenstates without normalization are then given by%
\begin{equation}
	\left\vert \psi _{j,s}\right\rangle =\sum_{k=1}^{N}\left( \sqrt{\frac{\cos
			\theta }{\cos \phi }}\right) ^{k}\sin \frac{kj\pi }{N+1}\left\vert
	k,s\right\rangle ,
\end{equation}%
with $s=\pm1$ and the left eigenstates are%
\begin{equation}
	\left\langle \chi _{j,s}\right\vert =\sum_{k=1}^{N}\left( \sqrt{\frac{\cos
			\phi }{\cos \theta }}\right) ^{k}\sin \frac{kj\pi }{N+1}\left\langle
	k,s\right\vert .
\end{equation}%
These eigenstates would be useful in higher-order corrections.

The energy degeneracy of the Zeeman coupling is thus lifted by the
perturbations. Consequently, we find that, if $\cos \theta \cos \phi >0$,
then the first-order corrections of the eigenenergies are real. Otherwise, if $%
\cos \theta \cos \phi <0$, the corrections are imaginary. 

Next, we study
the second-order corrections of the eigenenergies. The inter-subspace contributions in the second-order corrections are real: 
\begin{equation}
	E_{j,s}^{\rm {inter}\left( 2\right) }=\sum_{j^{\prime }}\frac{\left\vert
		\left\langle j^{\prime },-s\right\vert H_{1}\left\vert j,s\right\rangle
		\right\vert ^{2}}{E_{j,s}-E_{j^{\prime },-s}}=\sum_{j^{\prime }}\frac{%
		\left\vert \left\langle j^{\prime },-\right\vert H_{1}\left\vert
		j,s\right\rangle \right\vert ^{2}}{2s\delta_x}\in \mathbf{R.}
\end{equation}%
This explains why, when $\cos \theta \cos \phi
<0$, the curve of the energy spectra on the complex plane is not a straight line parallel to the imaginary axis.
Higher-order corrections contribute to the real part of the eigenenergies
when $\cos \theta \cos \phi <0$. 

We can then prove that, when $\cos \theta \cos \phi >0,$ all order corrections
of energies are real. In an arbitrary order correction, the denominators,
which are the product of factors such as $\left( E_{j,s}-E_{j^{\prime
	},-s}\right) $ or $\left( E_{j,s}+E_{j,s}^{\left( 1\right) }-E_{j^{\prime
	},s}-E_{j',s}^{\left( 1\right) }\right) $, are real. In the numerators, only terms such as%
\begin{equation}
	\left\langle j,s\right\vert H_{1}\left\vert j_{1},s_{1}\right\rangle
	\left\langle j_{1},s_{1}\right\vert H_{1}\left\vert j_{2},s_{2}\right\rangle
	\ldots \left\langle j_{m},s_{m}\right\vert H_{1}\left\vert j,s\right\rangle 
	\label{numerator}
\end{equation}%
exist. The term in Eq. (\ref{numerator}) is a product of $m+1$ matrix elements. Based on the form of $H_{1}$, we have the following
properties: 
\begin{eqnarray}
	\left\langle j_{k},s\right\vert H_{1}\left\vert j_{k^{\prime
	}},s\right\rangle  &=&\cos \theta \text{ or } \cos \phi \text{ or } 0, \\
	\left\langle j_{k},+\right\vert H_{1}\left\vert j_{k^{\prime
	}},-\right\rangle  &=&i\sin \theta \text{ or } i\sin \phi \text{ or } 0.
\end{eqnarray}%
Since the product in Eq. (\ref{numerator}) starts from and ends to the same state with spin $s$, there is always an even number of purely imaginary matrix elements in Eq. (\ref{numerator}). Thus, the numerators which consist of terms as in Eq. (\ref{numerator}) are real. To
sum up, when $\cos \theta \cos \phi >0,$ arbitrary order corrections are real, and the system exhibits $\mathcal{PT}$ symmetry. When $\cos \theta \cos \phi <0$, the system is in anti-$\mathcal{PT}$ symmetric phase.

\subsection{Symmetry}
Even though the essential dynamics for systems in anti-$\mathcal{PT}$ symmetric phases might be remarkably different from their $\mathcal{PT}$ symmetric counterparts, the anti-$\mathcal{PT}$-symmetric Hamiltonian $H^{\mathcal{APT}}$ can be conveniently defined in terms of a $\mathcal{PT}$-symmetric Hamiltonian $H^{\mathcal{PT}}$, such that $H^{\mathcal{APT}}=\pm iH^{\mathcal{PT}}$ \cite{Choi2018}. The Hermitian Zeeman coupling can be neglected when we consider the symmetry of the Hamiltonian in our non-Hermitian systems. We assume that we have a Hamiltonian for ${\mathcal{PT}}$ symmetry as follows    
\begin{equation}
	H^{\mathcal{PT}}=\left( 
	\begin{array}{cc}
		e^{i\theta }e^{iq}+e^{i\phi }e^{-i{q}} & 0 \\ 
		0 & e^{-i\theta }e^{i{q}}+e^{-i\phi }e^{-i{q}}%
	\end{array}
	\right) .
\end{equation}
In contrast, the system is in the anti-${\mathcal{PT}}$ symmetric phase for $\cos \theta \cos \phi <0$.  In other words, a Hamiltonian for the anti-${\mathcal{PT}}$ symmetry can be obtained when the phase $\phi$ shifts to $\phi + \pi$. We then have 
\begin{equation}
	H^{\mathcal{APT}}=\left( 
	\begin{array}{cc}
		e^{i\theta }e^{iq}-e^{i\phi }e^{-i{q}} & 0 \\ 
		0 & e^{-i\theta }e^{iq}-e^{-i\phi}e^{-i{q}}%
	\end{array}
	\right). 
\end{equation}
When the magnetic field is neglected, the PBC spectra of the Hamiltonian collapse into arcs and the wave vector $q$ is within the conventional Bloch Brillouin zone instead of the non-Bloch Brillouin zone. The band structure for $H^{\mathcal{APT}}$ is the same when $q \to q \pm \frac{\pi}{2}$. Thus, we have 
\begin{equation}
	H^{\mathcal{APT}}=\left( 
	\begin{array}{cc}
		\pm ie^{i\theta }e^{iq}\pm ie^{i\phi }e^{-iq} & 0 \\ 
		0 & \pm ie^{-i\theta }e^{iq} \pm ie^{-i\phi}e^{-iq}%
	\end{array}
	\right) =\pm iH^{\mathcal{PT}}.
\end{equation}
Note that the mathematical distinction between ${\mathcal{PT}}$ and anti-${\mathcal{PT}}$ systems is semantic in the framework of the pseudo-Hermitian quantum mechanics \cite{Choi2018}. Eq. (\ref{PH}) demonstrates that our system indeed respects the pseudo-Hermiticity when hopping phases are real. 

\subsection{GBZ of the system with $\mathcal{PT}$ symmetry}
\begin{figure}[H]
	\centering
	\includegraphics[width=\linewidth]{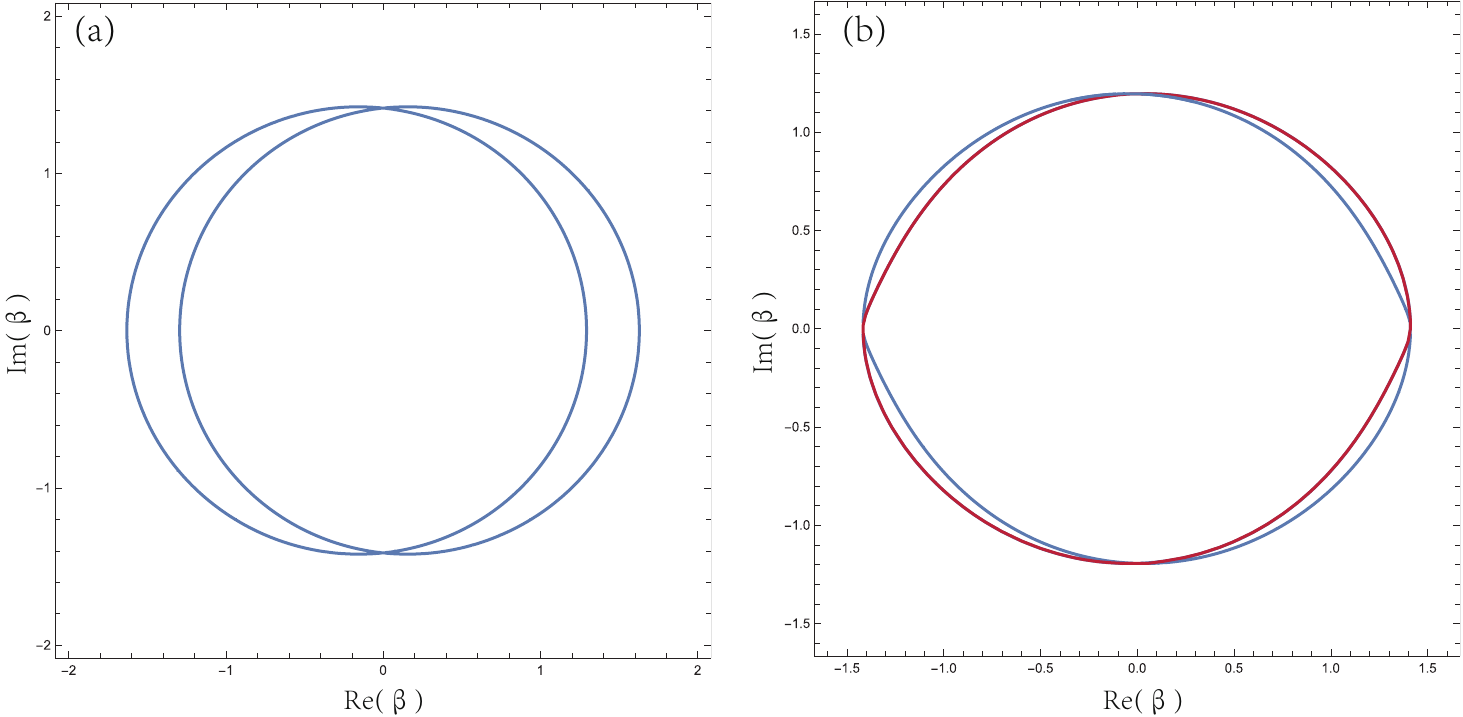}
	\caption{The GBZs for (a) the $\mathcal{PT}$ phase with $\theta=\pi/3,\phi=0,\delta_x=10$, and (b) the $\mathcal{APT}$ phase with $\theta=2\pi/3,\phi=0,\delta_x=9/4$. Two different colors are used in (b) to distinguish the two GBZs.} 
	\label{figs2}
\end{figure}

Here, we show the GBZ of the system with $\mathcal{PT}$ symmetry to compare with the GBZ of the system shown in Fig. 4(b). At the transition point $\cos\theta=0$ or $\cos\phi=0$, each of the two-branch energy spectra has three sub-branches, corresponding to three cusps in the GBZ. In the case that all eigenenergies are real ($\cos\theta \cos\phi >0$), one example GBZ is plotted in Fig. \ref{figs2}(a), and one example GBZ of the $\mathcal{APT}$ phase is shown in Fig. \ref{figs2}(b).
We note that the two circles of GBZ are too close to distinguish under a strong magnetic field for the $\mathcal{APT}$ phase.

\subsection{Signature of the non-Bloch $\mathcal{PT}$ symmetry}

When $\mathcal{PT}$ symmetry occurs, and all eigenenergies are real, we find a unique signature of the system, $\widetilde{\sigma }_{n,z}^{(j)}=0$ for an arbitrary state $n$ at any site $j$. The proof is given as follows. The eigenvalue equations in this case are given by Eqs. (\ref{en1})-(\ref{en3}). Given that $E_n \in \mathbf{R}$, the complex conjugate of these equations gives the relation 
\begin{equation}
	x^n_{j,-}=\left( x^n_{j,+}\right)^* \cdot e^{i \gamma},
	\label{x1}
\end{equation}
where $\gamma$ is a constant. Based on this relation, we consequently have 
\begin{equation}
	\widetilde{\sigma }_{n,z}^{(j)}= \left|x^n_{j,+}\right|^2-\left|x^n_{j,-}\right|^2=0.
	\label{x2}
\end{equation}
which can be verified numerically.

\begin{figure}[H]
	\centering
	\includegraphics[width=0.5\linewidth]{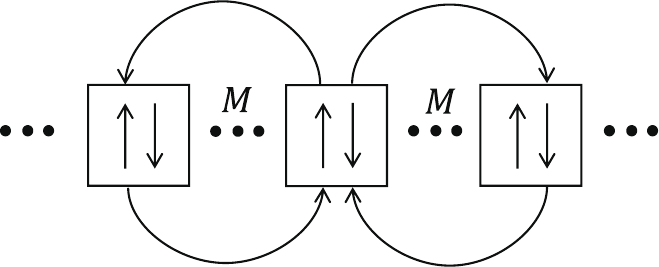}
	\caption{Schematic of a 1-D general system. A unit cell includes spin degrees of freedom, and the range of hopping is $M$.} 
	\label{figs3}
\end{figure}

In fact, the signature of $\widetilde{\sigma }_z=0$ can be generalized to an arbitrary 1-D open chain, as shown in Fig. \ref{figs3}. A unit cell is composed of spin up and spin down states, and the range of hopping is $M$, meaning that electrons hop up to the $M$-th nearest unit cells. Again, the $n$-th eigenvalue equation is
$H\psi _{n} =E_{n}\psi _{n},$ and the following eigenvalue equations are satisfied:
\begin{eqnarray*}
	\delta_x x_{1,-s}^{n}+e^{is\theta_1}x_{2,s}^{n}+e^{is\theta_2}x_{3,s}^{n}+\ldots +e^{is\theta_M}x_{M+1,s}^{n} &=&E_{n}x_{1,s}^{n}, \\
	e^{is\phi_1}x_{1,s}^{n}+\delta_x x_{2,-s}^{n}+e^{is\theta_1}x_{3,s}^{n}+e^{is\theta_2}x_{4,s}^{n}+\ldots +e^{is\theta_M}x_{M+2,s}^{n} &=&E_{n}x_{2,s}^{n}, \\
	e^{is\phi_M}x_{j-M,s}^{n}+e^{is\phi_{M-1}}x_{j-(M-1),s}^{n}+\ldots+\delta_x x_{j,-s}^{n}+\ldots+e^{is\theta_{M-1}}x_{j+M-1,s}^{n}+e^{is\theta_M}x_{j+M,s}^{n} &=&E_{n}x_{j,s}^{n}, \\
	e^{is\phi_M}x_{N-(M+1),s}^{n}+e^{is\phi_{M-1}}x_{N-M,s}^{n}+\ldots+\delta_x x_{N-1,-s}^{n}+e^{is\theta_1}x_{N,s}^{n} &=&E_{n}x_{N-1,s}^{n}, \\
	e^{is\phi_M}x_{N-M,s}^{n}+e^{is\phi_{M-1}}x_{N-(M-1),s}^{n}+\ldots+\delta_x x_{N,-s}^{n} &=&E_{n}x_{N,s}^{n}, 
\end{eqnarray*}
with $j\in \left[M+1,N-M\right]$ and $s=\pm 1$. $\theta_k$ and $\phi_k$ are hopping phases to the $k$-th nearest unit cell with $k\le M$. Assuming $E_n \in \mathbf{R}$, the complex conjugate of these equations also leads to Eqs. (\ref{x1}) and (\ref{x2}). 

\begin{figure}[H]
	\centering
	\includegraphics[width=1\linewidth]{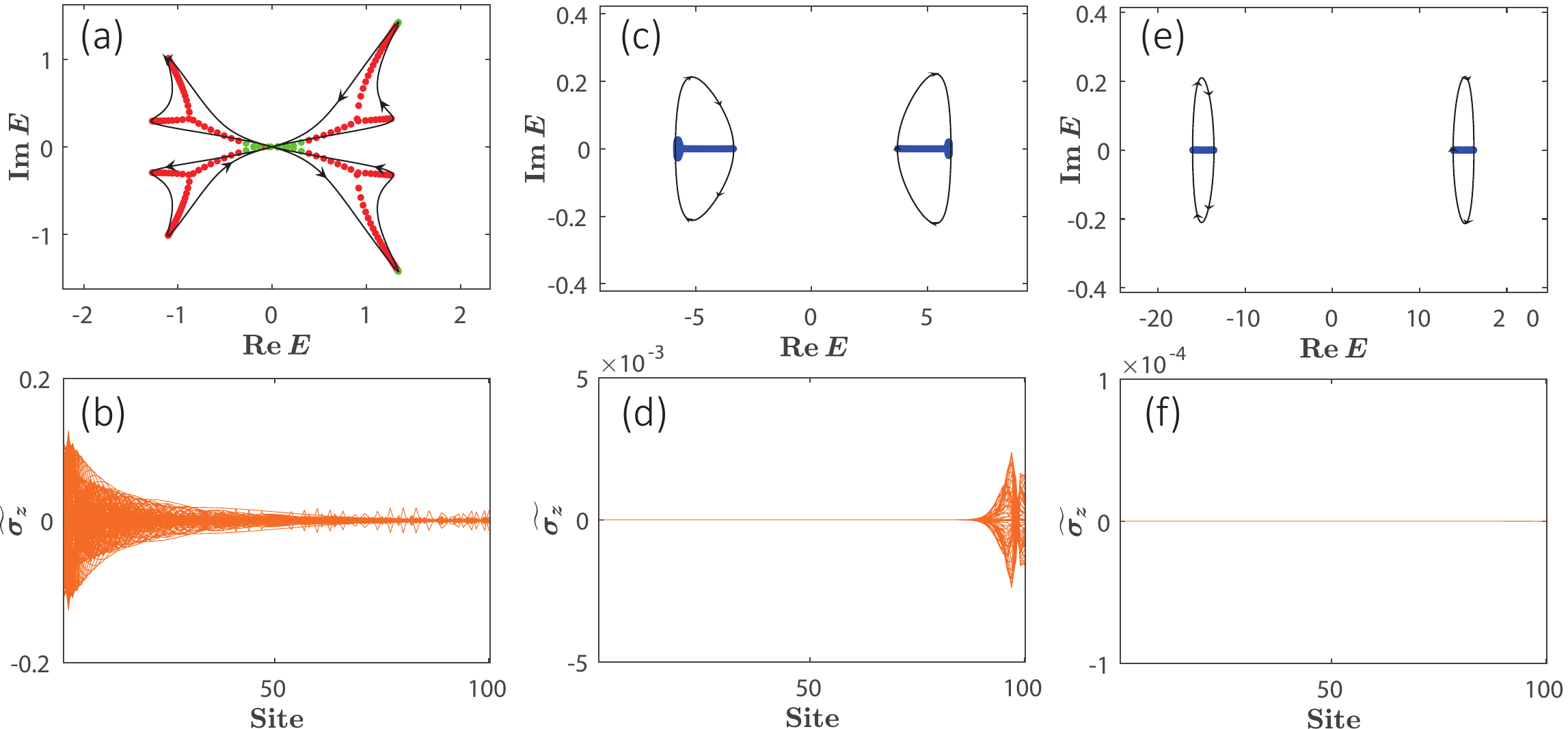}
	\caption{(a), (c), (e) Evolution of energy spectra with increasing magnetic field amplitude. The system parameters are $\theta_1=\frac{\pi}{3}, \phi_1=\frac{\pi}{4}$, $\theta_2=0.1 \theta_1, \phi_2=0.1 \phi_1$, and $\delta_x=1, 5, 15$ for (a), (c), (e), respectively. (b), (d), (f) Evolution of the spin component $\tilde{\sigma}_{z}$. The parameters in (b), (d), (f) are the same as in (a), (c), (e), respectively.} 
	\label{figs4}
\end{figure}

Here, we show a concrete example including next-nearest hopping in Fig. \ref{figs4}. With increasing $\delta_x$, 
the non-Hermiticity decreases and the OBC spectra change from partially to entirely real [Figs. \ref{figs4}(a), \ref{figs4}(c), and \ref{figs4}(e)]. Meanwhile, the corresponding spin polarizations of $\tilde{\sigma}_z$ with increasing magnetic field $\delta_x$ are shown in Figs. \ref{figs4}(b), \ref{figs4}(d), and \ref{figs4}(f). $\tilde{\sigma}_z=0$ indeed corresponds to the system under the exact non-Bloch $\mathcal{PT}$ phase with all eigenvalues real in OBC.

\section{Details of eigenenergies and their skin direction in the 2D Model}

We generalize the model to two dimensions, and the Hamiltonian is given by
\begin{eqnarray}
	H & =&\sum_{i,j}c_{i,j}^{\dag} e^{i\theta \sigma _{z}}c_{i+1,j}
	+c_{i+1,j}^{\dag} e^{i\phi \sigma _{z}}c_{i,j}  \label{eq_2D} \\ 
	&+& c_{i,j}^{\dag}e^{i \theta' \sigma _{z}}c_{i,j+1} 
	+c_{i,j+1}^{\dag}e^{i\phi' \sigma _{z}}c_{i,j}
	+c_{i,j}^{\dag} (\boldsymbol{\delta \cdot \sigma})c_{i,j}, \notag 
\end{eqnarray}
where $c_{i,j}$ and $c_{i,j}^{\dag}$ are the annihilation and creation operators, respectively, at site $\left(i,j\right) $ with spin, and the phases $\theta,\phi,\theta',\phi'$ are tunable parameters. The Zeeman term $\boldsymbol{\delta \cdot \sigma}$ can be induced by a magnetic background \cite{kato2019}, or an in-plane magnetic field where the gauge can be chosen to vanish. When a real in-plane magnetic field is applied, the gauge can be chosen as $\mathbf{A}=(0,0,\delta_x y - \delta_y x)$ or $(0,-\delta_x z,0)$. The lattice model could be gauge free. Thus, the eigenenergies are gauge-invariant, and the skin effects remain in different gauge.

\begin{figure}[H]
	\centering
	\includegraphics[width=0.8\linewidth]{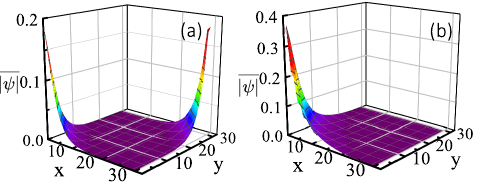}
	\caption{The 2D NHSE for a system with $\theta=\theta'=\frac{i\pi}{4}, \phi=\phi'=\frac{\pi}{3}$ (a) without and (b) with Zeeman coupling $\boldsymbol{\delta}=(5,0,0)$. For simplicity, the plots show the average of the amplitudes of the wave functions, $\overline{|\psi|} = \sum_i |\psi_i|$. The skin effect is cornered in the presence of a magnetic field.} 
	\label{figs5}
\end{figure}

As shown in Fig. \ref{figs5}(a), when excluding Zeeman coupling, the spin-up states are concentrated at the left-down ($-x$ and $-y$ direction) corner, while the spin-down states are localized at the right-up corner in an open lattice, which is a typical higher-order skin effect \cite{Fu2021}. When gradually increasing for simlpicity the Zeeman coupling $\delta_x \sigma_x$, all OBC states tend to skin to a specific corner. We therefore conclude that we can control the system parameters to localize states to any corner [Fig. \ref{figs5}(b)]. 

\begin{figure}[H]
	\centering
	\includegraphics[width=1\linewidth]{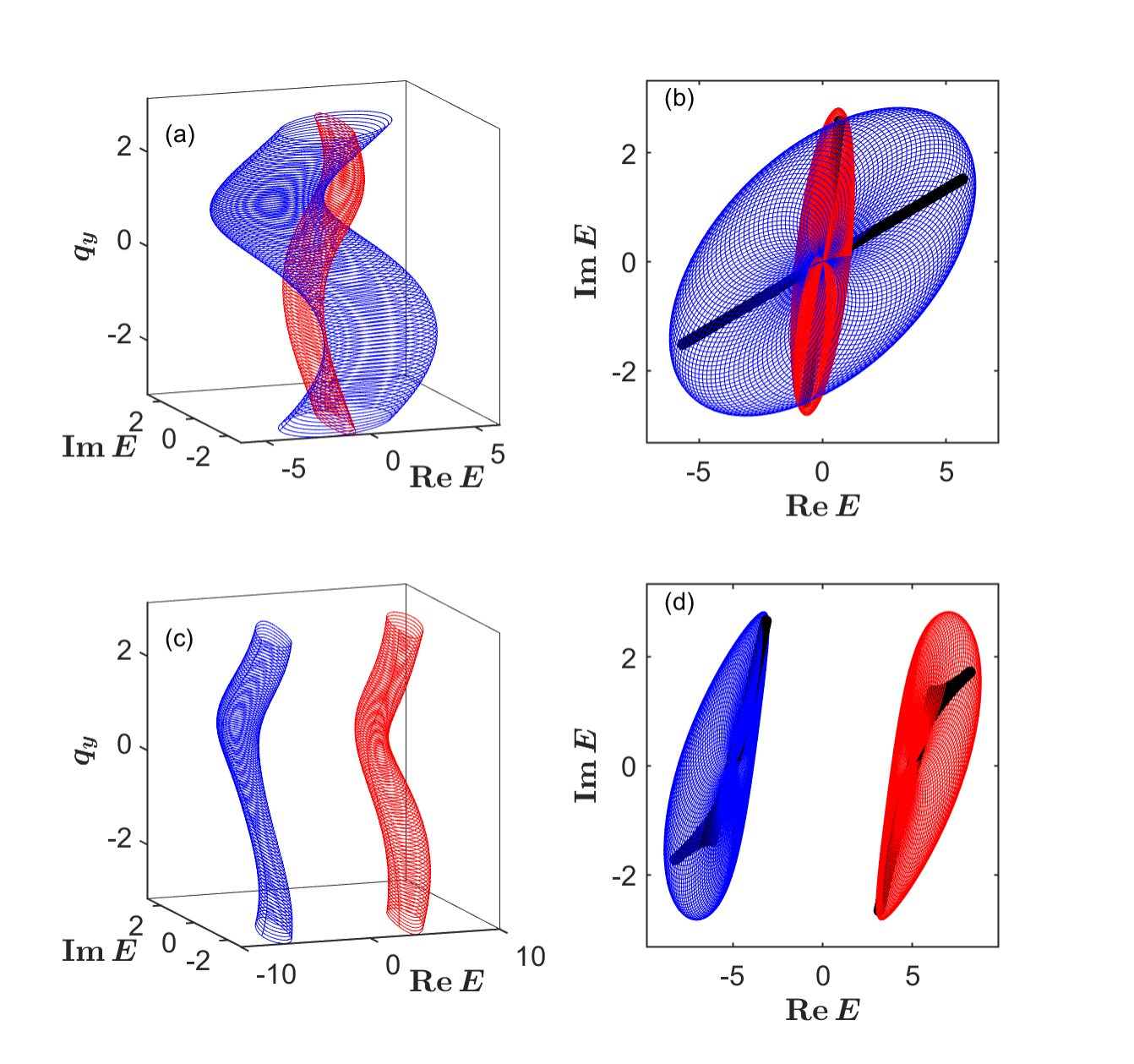}
	\caption{The energy spectra of the 2D model with the same phase parameters as in Fig. 5 in the main text, $\theta=\theta'=\frac{i\pi}{4},\phi=\phi'=\frac{\pi}{3}$. (a) The $q_y-$slice energy spectra and (b) the energy curves projected to the $\text{Re}E-O-\text{Im}E$ plane in the absence of magnetic field. (c) The $q_y-$slice energy spectra and (d) the energy curves projected to the $\text{Re}E-O-\text{Im}E$ plane with $\delta_x=5$.} 
	\label{figs6}
\end{figure}

To show the complex energy spectra in two-dimensional momentum space, we first fix $q_y$ and plot the energy curves with varying $q_x$ in a slice. We then stack these curves to form a $q_y-$slice plot, as shown in Figs. \ref{figs6}(a) and \ref{figs6}(c). 
When all the energy curves are projected to the complex plane $\text{Re}E-O-\text{Im}E$, the PBC energy curves form areas which cover all the eigenenergies in OBC (black dots), as shown in Figs. \ref{figs6}(b) and \ref{figs6}(d). 
In Figs. \ref{figs6}(a) and \ref{figs6}(b) we show the energy spectra with parameters $\theta=\theta'=\frac{i\pi}{4},\phi=\phi'=\frac{\pi}{3}$, without magnetic field.
Similar to the 1D model, the PBC spectra for two spins without Zeeman coupling are also distributed in two independent Riemann surfaces, as shown in Fig. \ref{figs6}(a). 

The curves with varying $q_x$ in the slices with fixed $q_y$ have winding number $\omega = \pm 1$, representing skinning to the left and right of the $x$-axis, respectively. In addition, the curves with varying $q_y$ in the slices with fixed $q_x$ have winding number $\omega = \pm 1$, representing skinning to the left and right of the $y$-axis, respectively.
In Figs. \ref{figs6}(c) and \ref{figs6}(d), we show the energy spectra with the same parameters as before $\theta=\theta'=\frac{i\pi}{4},\phi=\phi'=\frac{\pi}{3}$, but with a magnetic field $\delta_x=5$.

The determination of the skin direction in the 2D model is in fact similar to the one in the 1D model. 
We can therefore control the skin effect to locate at any two corners or one corner on the 2D lattice, by tuning the phase parameters $(\theta,\phi,\theta',\phi')$ or the Zeeman coupling. 
The eigenstates plotted in Figs. \ref{figs6}(b) and \ref{figs6}(d) have skin modes shown in Figs. \ref{figs5}(a) and \ref{figs5}(b), respectively.

\end{document}